\documentclass[preprint]{aastex}

\usepackage{natbib}

\slugcomment{prepared for the  Astronomical Journal, 2003}

\def\bn{\bigskip\noindent}

\def\h{{\rm h}}
\def\m{{\rm m}}
\def\s{{\rm s}}
\def\iner{{\rm iner}}

\def\OO#1{{\cal O}(c^{-#1})}
\def\ve#1{{\bf #1}}

\def\vD{\ve{D}}

\def\vr{\ve{r}}
\def\vw{\ve{w}}
\def\vx{\ve{x}}
\def\vv{\ve{v}}

\def\vS{\ve{S}}

\def\vQ{\ve{Q}}
\def\vX{\ve{X}}

\def\vOmega{\ve{\Omega}}
\def\vSigma{\ve{\Sigma}}

\def\cO{{\cal O}}
\def\cM{{\cal M}}
\def\cS{{\cal S}}
\def\C{{\cal C}}
\def\S{{\cal S}}

\def\beqn{\begin{eqnarray*}}
\def\eeqn{\end{eqnarray*}}

\shorttitle{Explanatory supplement for the IAU 2000 Resolutions on Relativity}
\shortauthors{M. Soffel et al.}

\begin{document}

\title{The IAU 2000 resolutions for astrometry,
celestial mechanics and metrology in the relativistic framework:
explanatory supplement}

\author{M.~Soffel\altaffilmark{1},
S.A.~Klioner\altaffilmark{1},
G.~Petit\altaffilmark{2},
P.~Wolf\altaffilmark{2},
S.M.~Kopeikin\altaffilmark{3},
P.~Bretagnon$^\dagger$\altaffilmark{4},
V.A.~Brumberg\altaffilmark{5},
N.~Capitaine\altaffilmark{6},
T.~Damour\altaffilmark{7},
T.~Fukushima\altaffilmark{8},
B.~Guinot\altaffilmark{6},
T.~Huang\altaffilmark{9},
L.~Lindegren\altaffilmark{10},
C.~Ma\altaffilmark{11},
K.~Nordtvedt\altaffilmark{12},
J.~Ries\altaffilmark{13},
P.K.~Seidelmann\altaffilmark{14},
D.~Vokrouhlick\'y\altaffilmark{15},
C.~Will\altaffilmark{16},
Ch.~Xu\altaffilmark{17}}

\altaffiltext{1}{Lohrmann Observatory, Dresden Technical University,
Mommsenstr. 13, 01062 Dresden, Germany}
\altaffiltext{2}{BIPM, Pavillon de Breteuil, 92312 Sevres, France}
\altaffiltext{3}{Department of Physics and Astronomy, University of Missouri at Columbia, 322 Physics Building, Columbia, MO 65211, USA}
\altaffiltext{4}{Bureau des longitudes, 75014 Paris, France}
\altaffiltext{5}{Institute of Applied Astronomy, Kutuzova emb. 10, 191187 St.Petersburg, Russia}
\altaffiltext{6}{Observatoire de Paris, 61 Avenue de l'Observatoire, 75014 Paris, France}
\altaffiltext{7}{Institut des Hautes Etudes Scientifiques, 35, route
de Chartres, F-91440 Bures sur Yvette,
 France}
\altaffiltext{8}{National Astronomical Observatory, 2-21-1 Ohsawa, Mitaka, Tokyo 181, Japan}
\altaffiltext{9}{Department of Astronomy, Nanjing University, National Astronomical Observatory, 210093 Nanjing, China}
\altaffiltext{10}{Lund Observatory, Box 43, S-22100 Lund, Sweden}
\altaffiltext{11}{NASA Goddard Space Flight Center, Code 926, Greenbelt, MD 20771, USA}
\altaffiltext{12}{e-mail: kennordtvedt@imt.net}
\altaffiltext{13}{Center for Space Research, Mail Code R1000, The University of Texas at Austin, Austin, Tx 78712, USA}
\altaffiltext{14}{Department of Astronomy, University of Virginia, PO Box 3818, Charlottesville, VA 22903, USA}
\altaffiltext{15}{Astronomical Institute, Charles University, Svedska 8, CZ-150 00 Praha 5, Czech Republic}
\altaffiltext{16}{McDonnell Center for the Space Sciences, Department of Physics, Washington University, St Louis, Missouri 63130, USA}
\altaffiltext{17}{Nanjing University, Nanjing, China}

\begin{abstract}
This paper discusses the IAU Resolutions B1.3, B1.4, B1.5 and B1.9
(2000) that were adopted during the 24th General Assembly in
Manchester, 2000 and provides details and explanations for these
Resolutions. It is explained why they present significant progress over
the corresponding IAU 1991 Resolutions and why they are necessary  in
the light of present accuracies in astrometry, celestial mechanics and
metrology. In fact most of these Resolutions are consistent with
astronomical models and software already in use.

The metric tensors and gravitational potentials of both the Barycentric
Celestial Reference System and Geocentric Celestial Reference System
are defined and discussed. The necessity and relevance of the two
celestial reference systems are explained. The transformations of
coordinates and gravitational potentials are discussed. Potential
coefficients parameterizing the post-Newtonian gravitational potentials
are expounded. Simplified versions of the time transformations suitable
for modern clock accuracies are elucidated. Various approximations used
in the Resolutions are explicated and justified. Some models (e.g. for
higher spin moments) that serve the purpose for estimating orders of
magnitude have actually never been published before.
\end{abstract}


\section{Introduction}

It is clear that, beyond some threshold of accuracy, any astronomical
problem has to be formulated within the framework of Einstein's theory
of gravity (General Relativity Theory, GRT). Many high precision
astronomical techniques have already passed this threshold.   For
example, Lunar Laser Ranging (LLR) measures the distance to the Moon
with a precision of a few cm,  thereby operating at the $10^{-10}$
level.  At this level several relativity effects are significant and
observable. Relativity effects related with the motion of the
Earth-Moon system about the Sun are of the order $(v_{\rm orbital}/c)^2
\simeq 10^{-8}$. The Lorentz contraction of the lunar orbit about the
Earth that appears in barycentric coordinates has an amplitude of about
$100\,$cm whereas in some suitably chosen (local) coordinate system
that moves with the Earth-Moon barycenter the dominant relativistic
range oscillation reduces to a few cm only
\citep{Mashhoon:1985,Soffel:et:al:1986}.

The situation is even more critical in the field of astrometry. It is
well known that the gravitational light deflection at the limb of the
Sun amounts to $1.75\arcsec$ and decreases only as $1/r$ with
increasing impact parameter $r$ of a light ray to the solar center.
Thus, for light rays incident at about $90^\circ$ from the Sun the
angle of light deflection still amounts to 4 mas. To describe the
accuracy of astrometric measurements it is useful to make use of  the
parameter $\gamma$ of the parameterized post-Newtonian (PPN) formalism.
We would like to emphasize that this paper deals solely with Einstein's
theory of gravity, where $\gamma=1$, and not with the PPN formalism. Nevertheless, the
introduction of $\gamma$ is useful if one talks about measuring
accuracies. In the PPN formalism, the angle of light deflection is
proportional to $(\gamma +
1)/2$ so that astrometric measurements might be used for a precise
determination of the parameter $\gamma$. Meanwhile, VLBI  has achieved
accuracies of better than 0.1 mas, and regular geodetic VLBI
measurements have frequently been used to determine the space curvature
parameter. A detailed analysis of VLBI data from the projects MERIT and
POLARIS/IRIS gave $\gamma = 1.000 \pm 0.003 \, ,$
\citep{Robertson:et:al:1984,Carter:et:al:1985} where a formal standard
error is given. Recently an advanced processing of VLBI data provided
the best current estimates $\gamma = 0.9996\pm 0.0017$
\citep{Lebach:et:al:1995} and $\gamma = 0.99994\pm 0.00031$
\citep{Eubanks:et:al:1997}. Current accuracy of modern optical
astrometry as represented by the HIPPARCOS catalog is about 1
milliarcsecond, which gave a determination of $\gamma$ at the level of
$0.997\pm0.003$  \citep{Froeschle:Mignard:Arenou:1997}. Future
astrometric missions such as SIM and especially GAIA will push the
accuracy to the level of a few  microarcseconds ($\mu$as), and the
expected accuracy of determinations of $\gamma$ will be $10^{-6} -
10^{-7}$. The accuracy of 1 $\mu$as should be compared with the maximal
possible light deflection due to various parts of the gravitation
field: the post-Newtonian effect of $1.75\arcsec$ due to the mass of
the Sun, $240 \, \mu$as caused by the oblateness of Jupiter,
$J_2$  ($10\, \mu$as due to Jupiter's $J_4$), the post-post-Newtonian
effect of $11\,\mu$as due to the Sun, etc. This illustrates how
complicated the relativistic modeling of future astrometric
observations will be. It is clear that for such high accuracy
the corresponding model must be formulated in a self-consistent
relativistic framework.

Another problem worth mentioning is that  of time measurement. The
realization of the SI second (the unit of proper time) has improved by
one order of magnitude in the last few years with the advent of laser
cooled atomic clocks \citep[][and references
therein]{Lemonde:et:al:2001,Weyers:et:al:2001}, and is now below two
parts in $10^{15}$. This should be compared with the dimensionless
quantity $U_E/c^2 \simeq 7 \times 10^{-10}$, which gives the
order of magnitude of relativistic effects produced by the gravity
field of the Earth itself in the vicinity of its surface. In the near
future, laser cooled atomic clocks in micro-gravity are expected to
lead to a further improvement by at least one order of magnitude. At
present several clock experiments in terrestrial orbit are planned,
such as the Atomic Clock Ensemble in Space project
\citep[ACES,][]{Lemonde:et:al:2001}. These in turn are likely to lead to
clock experiments in solar orbits like the Solar Orbit Relativity
Test (SORT) project. All of these experiments require a detailed account of
many subtle relativistic effects.

Finally we would like to mention the problem of geodesic
precession-nutation \citep[a relativistic effect that
is discussed in more detail below,][]{MTW:1973,Soffel:1989} and the description of
Earth's rotation in a suitably chosen Geocentric Celestial Reference
System (GCRS). Geodesic precession amounts to 1.9''/century and
geodesic nutation is dominated by an annual term with amplitude 0.15~mas.
Since the GCRS is chosen to be kinematically
non-rotating geodesic precession-nutation
should be contained in the model
describing the relation between the GCRS and the International
Terrestrial Reference System (ITRS). According to the IAU 2000
Resolution B1.6 this relativistic precession-nutation is indeed
contained in the present IAU precession-nutation model.

These examples show clearly that high-precision modern astronomical
observations  can no longer be described by Newtonian theory,  but
require Einstein's theory of gravity.

The consequences of this are profound for the basic formalism to be
used since one often tends to express it in terms of ``small
relativistic corrections'' to Newtonian theory. This can lead to
misconceptions and mistakes. One central point is that in Newton's
theory, globally preferred coordinate systems exist that have a direct
physical meaning. In the Newtonian framework, idealized clocks show
absolute time everywhere in the universe at all times, and global
spatial inertial coordinates exist in which  dynamical equations of
motion show no inertial forces.  This is no longer true in GRT. Usually
space-time coordinates have no direct physical meaning and it is
essential to construct the observables as coordinate independent
quantities, i.e., scalars, in mathematical language. This construction
usually occurs in two steps: first one formulates a coordinate picture
of the measurement procedure and then one derives the observable out of
it. This leads us to the problem of defining useful and adequate
coordinate systems in astronomy.  The underlying concept in
relativistic modeling of astronomical observations is a relativistic
four-dimensional {\sl reference system} (RS).  By {\sl reference
system} we mean a purely mathematical construction (a chart or a
coordinate system) giving ``names'' to space-time events.  In contrast to
this a {\sl reference frame} is some materialization of a reference
system. In astronomy the materialization is usually realized by means
of a catalog or ephemeris, containing positions of some celestial
objects relative to the reference system under consideration. Hence it
is very important to understand  that any reference frame is defined
only through a well-defined reference system, which has been used to
construct physical models of observations.

In the following, a 4-dimensional space-time reference system will be
described by four coordinates $x^\alpha = (x^0,x^i)=(x^0,x^1,x^2,x^3)$.
Here and below the greek indices (e.g., $\alpha$) take the values
$0,1,2,3$ and the latin ones (e.g., $i$) take the values $1,2,3$. The
index $0$ refers to the time variable and the indices $1,2,3$ refer to
the three spatial coordinates. For dimensional reasons one usually
writes $x^0 = c\,t$ where $c$ is the speed of light and $t$ a time
variable. According to the mathematical formalism of General
Relativity, a particular reference system is fixed by the specific form
of the  metric tensor $g_{\alpha\beta}(t,x^i)$, which allows one to
compute the 4-distance $ds$ between any two events $x^\alpha$ and
$x^\alpha+dx^\alpha$ according to the rule

\begin{eqnarray}\label{ds2}
ds^2=g_{\alpha\beta}(t,x^i)\,dx^\alpha\,dx^\beta \equiv
g_{00} c^2 dt^2 + 2 g_{0i} c\, dt\,  dx^i + g_{ij} dx^i dx^j \, ,
\end{eqnarray}

\noindent
where Einstein's summation convention (summation over two equal
indices) is implied. The  metric tensor allows one to
derive translational and rotational equations of motion of bodies, to
describe the propagation of light, and to model the process of
observation. Examples of such modeling include relating the observed
(proper) time of an observer to the coordinate time $t$, or relating
the angles between two incident light rays as observed by that observer
to the corresponding coordinate directions. All of these components can
be combined into a single relativistic model for a particular kind of
observations. Such a model contains a certain set of parameters
describing various properties of the objects participating in the
process of observations.  These parameters should be determined from
observations. Many of these parameters crucially depend on the
reference system used to formulate the model of observations (e.g., the
initial positions and velocities of certain bodies). Some other
parameters might not depend at all on the reference system (e.g., the
speed of light in vacuum). On the other hand, according to the
principle of covariance different reference systems covering the region
of space-time under consideration are mathematically equivalent in the
sense that any such system can be used to model the observations.  This
freedom to choose the reference system can be used to simplify the
models or to make the resulting parameters more physically adequate.

It is widely accepted  that in order to describe adequately modern
astronomical observations one has to use several relativistic reference
systems. The solar system Barycentric Celestial Reference System (BCRS)
can be used to model light propagation from distant celestial
objects as well as the motion of bodies within the solar system.  The
Geocentric Celestial Reference System is physically adequate to
describe processes occurring in the vicinity of  the Earth (Earth's
rotation, motion of Earth's satellites etc.). The introduction of
further local systems (selenocentric, martian etc.) might be of
relevance for specific applications, where physical phenomena in the vecinity of the
corresponding body play a role.

The necessity to use several reference systems can be understood from
the following example. If we were to characterize terrestrial observers
by the difference between their BCRS coordinates and the BCRS
coordinates of the geocenter, the positions of the observers relative
to the geocenter would be altered by time-dependent, relativistic
coordinate effects (such as Lorentz contraction) which have nothing to
do with the Earth's rotation or with geophysical factors and which
would vanish if one employs suitable GCRS coordinates instead.
On the other hand, the coordinate positions derived with VLBI
observations are used to investigate local geophysical processes  and
some adequate geocentric reference system allows one to simplify their
description.

The basic idea is to construct a special local reference system for
each material subsystem, where relativistic equations of motion of a
test body inside the considered subsystem take a particularly simple
form. In such a local reference system the influence of external
matter, in accordance with the equivalence principle, should be given
by tidal potentials only, that is, by potentials whose expansions in
powers of local spatial coordinates in the vicinity of the origin of
the corresponding reference system starts with the second order (the
linear terms representing inertial forces may also exist, but can be
eliminated if desired by a suitable choice of the origin of the local
coordinates).

Two advanced relativistic formalisms have been worked out to tackle
this problem in the first post-Newtonian approximation of General
Relativity.  One formalism is due to Brumberg and Kopeikin
\citep{Brumberg:Kopeikin:1989,Kopeikin:1988,Kopeikin:1990,Brumberg:1991,Klioner:Voinov:1993}
and another one is due to Damour, Soffel and Xu
\citep{Damour:Soffel:Xu:1991,Damour:Soffel:Xu:1992,Damour:Soffel:Xu:1993,Damour:Soffel:Xu:1994},
frequently called the DSX framework. The IAU 2000 Resolutions B1.3-B1.5
are based on  these approaches. These Resolutions extend corresponding
older ones that are reconsidered in the next section. From a
mathematical point of view Resolution B1.3 recommends the use of
certain coordinates and the way of writing the metric tensor. Clearly
one  might use any coordinate system that might be well adapted to the
specific problem of interest. Nevertheless because of the high risk of
possible confusion the strategy of recommending special coordinate
systems (to fix the gauge completely in the mathematical language) has
significant advantages. If different coordinates are employed to derive
certain results this should be stressed explicitly so that they can be
transformed into the reference systems recommended by the IAU and can
be compared with the results derived in the IAU framework.

\begin{figure}
\epsscale{1.0}
\plotone{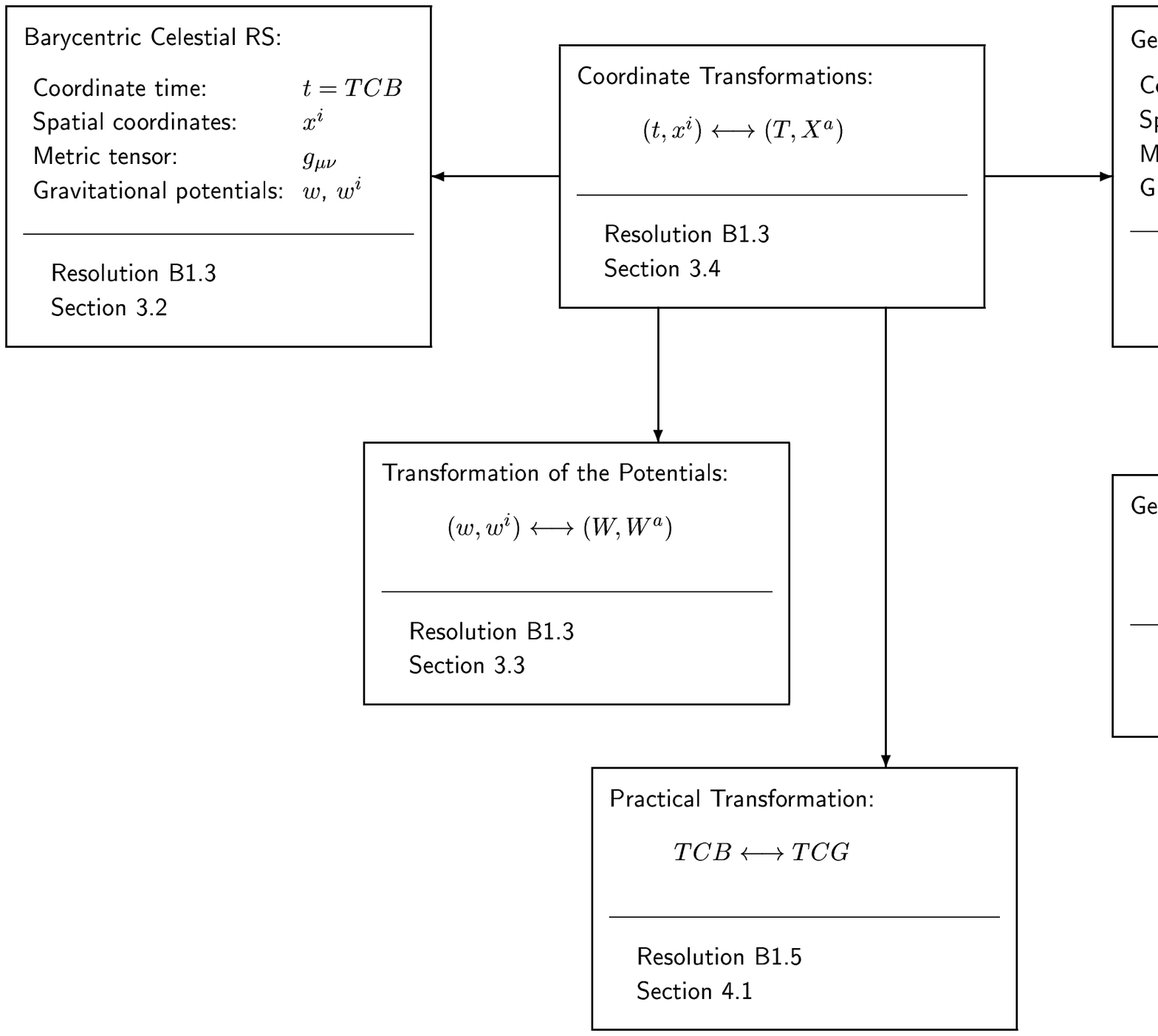}
\caption{
Notations used for quantities of the Barycentric
and Geocentric Celestial Reference Systems
(coordinates, metric, potentials, and multipole moments)
with references to the Sections and Resolutions where they appear.
\label{Figure-BCRS-GCRS-trasnformations}
}
\end{figure}

The organisation of the present paper is as follows (see also Figure
\ref{Figure-BCRS-GCRS-trasnformations}). In Section
\ref{Section-IAU1991-and-others} the principal content of the IAU 1991
recommendations on relativity (Section \ref{Section-IAU-1991}) and the
further related IAU and IUGG resolutions (Section
\ref{Section-1991-1997}) are repeated. The IAU 2000 resolutions on
relativity (Resolutions B1.3, B1.4, B1.5 and B1.9) are discussed in
Section \ref{Section-IAU-2000}. The full text of the IAU 2000
resolutions on relativity is given in Appendix \ref{Appendix-IAU-2000}.
Section \ref{Section-two-CRS} briefly clarifies the necessity and the
role of the two celestial reference systems defined by the IAU
resolutions. The Barycentric Celestial Reference System (BCRS) defined
by Resolution B1.3 is discussed in Section~\ref{Section-BCRS}. Section
\ref{Section-GCRS} is devoted to a discussion of the Geocentric
Celestial Reference System (GCRS) as well as the definition of the
geocentric gravitational potentials also defined by Resolution B1.3.
The coordinate transformations between the BCRS and GCRS, also fixed by
Resolution B1.3, are explained in Section \ref{Section-trans}.
Potential coefficients which can be used to represent in a meaningful
way the post-Newtonian geocentric gravitational potential of the Earth
in its immediate vicinity are fixed by Resolution B1.4 and explained in
Section \ref{Section-ML-SL}. As an illustration the gravitational
potentials of the BCRS are calculated in Section
\ref{Section-mass-monopoles} for the simplified case when all
gravitating bodies of the solar system can be characterized by their
masses only (no further structure of the gravitational field of the
bodies is considered). A similar form of the barycentric gravitational
potentials is used in Resolution B1.5 where a practical relativistic
framework for time and frequency applications in the solar system is
formulated. This practical relativistic framework is discussed in
Section \ref{Section-time-frequency}. The practical transformation
between the coordinate times of the BCRS and GCRS is explained in
Section \ref{Section-time-BCRS} while the transformations between
various kinds of the time scales appropriate for the Earth's vicinity
are discussed in Section \ref{Section-time-GCRS}. Appendix
\ref{Appendix-IAU-PPN} contains an explicit proof that the BCRS metric
coincides with well-known results from the literature.

\section {The IAU 1991 framework and  previous recommendations}
\label{Section-IAU1991-and-others}

\subsection{The IAU 1991 recommendations}
\label{Section-IAU-1991}

The IAU Resolution A4 (1991) contains nine recommendations, the first
five of which are directly relevant to our discussion.

In the first recommendation, the metric tensor for space-time
coordinate systems $(t,{\bf x})$ centered at the barycenter of an
ensemble of masses is recommended in the form

\begin{eqnarray}\label{iau_metric}
g_{00} &=& -1
+ {2U(t,{\bf x}) \over c^{2}} + \OO4,\nonumber\\
\nonumber\\
g_{0i} &=& \ \OO3,\\
\nonumber\\
g_{ij} &=& \delta_{ij}\,\left(1+{2U(t,{\bf x}) \over c^{2}}\right) +
\OO4 \, ,
\nonumber
\end{eqnarray}

\noindent
where $c$ is  the speed of light in vacuum ($c = 299792458$ m/s) and
$U$ is the Newtonian gravitational potential (here a sum of the
gravitational potentials of the ensemble of masses, and of an external
potential generated by bodies external to the ensemble, the latter
potential vanishing at the origin). The algebraic sign of $U$ is taken
to be  positive and it satisfies the Poisson equation

\begin{equation}\label{poisson}
\Delta U = - 4 \pi G \rho \, .
\end{equation}

\noindent
Here, $G$ is the gravitational constant, $\rho$ is the matter
density and $\Delta$ is the usual Laplacian $\Delta \equiv
\partial^2/\partial x^2 + \partial^2/\partial y^2 +
\partial^2/\partial z^2$, $\ve{x}=(x,y,z)$.
 This recommendation recognizes also that space-time cannot be
described by a single coordinate system. The recommended form of the
metric tensor can be used not only to describe the barycentric
reference system of the whole solar system, but also to define the
geocentric reference system centered in the center of mass of the Earth
with a suitable function  instead of $U$, now depending upon geocentric
coordinates. In analogy to the geocentric reference system a
corresponding reference system can be constructed for any other body of
the solar system.

In the second recommendation, the origin and orientation of the spatial
coordinate grids for the barycentric and geocentric reference systems
are defined. Notably it is specified that the spatial coordinates of
these systems should show no global rotation with respect to a set of
distant extragalactic objects. It also specifies that the SI
(International System of Units) second and the SI meter should be the
units of time and length in all coordinate systems. It states in
addition that the time coordinates should be derived from an Earth
atomic time scale.

The third recommendation defines $TCB$ (Barycentric Coordinate Time)
and $TCG$ (Geocentric Coordinate Time) as the time coordinates of the
BCRS and GCRS, respectively. Here we write $(t=TCB,x^i)$ and
$(T=TCG,X^i)$ for the respective coordinates. The recommendation also
defines the origin of the times scales in terms of International Atomic Time ($TAI$).
The reading of the coordinate time scales on 1977 January
1, $0^\h \, 0^\m \, 0^\s$ $TAI$ ($ JD = 2443144.5 \, TAI$) must be 1977
January 1, $0^\h \, 0^\m \, 32.184^\s$). Finally, the recommendation
declares that the units of
measurements of the coordinate times of all reference systems are
chosen so that they are consistent with the SI second.  The
relationship between $TCB$ and $TCG$ is then given by the time part of
the full 4-dimensional transformation between the barycentric and
geocentric reference systems

\begin{equation}\label{oldtcb-tcg}
TCB - TCG = c^{-2}\,
\left[\int^{t}_{t_0}
\left({v_{E}^{2} \over 2}
      + U_{\rm ext}(\vx_E) \right) \, dt
+ v^i_E r^i_E \right] + \OO4,
\end{equation}

\noindent
where $x^i_{E}$ and $v^i_{E}$ are the barycentric coordinate position
and velocity of the geocenter, $r^i_E = x^i - x^i_{E}$ with $x^i$ the
barycentric position of some observer, and $U_{\rm ext}(\ve{x}_E)$ is
the Newtonian potential of all solar system bodies apart from the Earth
evaluated at the geocenter.

In the fourth recommendation another coordinate time, Terrestrial Time
($TT$), is defined. It differs from $TCG$ by a constant rate only

\begin{equation}\label{TCG-TT}
TCG-TT=L_G\times(JD-2443144.5)\times 86400, \qquad
L_G \approx 6.969291\times 10^{-10},
\end{equation}

\noindent
where $JD$ is $TAI$ measured in Julian days, so that the mean rate of
$TT$ agrees with that of the proper time of an observer situated on the
geoid up to a certain accuracy limit. Up to a constant shift of 32.184 s,
$TT$ represents an ideal form of
$TAI$, the divergence between them being a consequence of the physical
defects of atomic clocks.  It is also recognized that  $TT$ is nothing
but a rescaling of the geocentric coordinate time $TCG$.

The fifth recommendation states that the old barycentric time $TDB$ may
still be used where discontinuity with previous work is deemed to be
undesirable. Let us note, however, that $TDB$ was never defined in a
self-consistent and exact manner. For that reason it cannot be used in
theoretical considerations. In the notes to the third recommendation
the relation of the $TCB$ with $TDB$ is given as

\begin{equation}\label{TCB-TDB}
TCB-TDB=L_B\times(JD-2443144.5)\times 86400, \qquad
L_B \approx 1.550505\times 10^{-8}.
\end{equation}

\noindent
Note, however, according to the IAU Resolution C7 (see Section
\ref{Section-1991-1997}) $JD$ is defined in Terrestrial Time ($TT$) which
makes this formula problematic.

\subsection{Further resolutions}
\label{Section-1991-1997}

Resolution 2  (1991)  of the International Union of Geodesy and
Geophysics (IUGG) defined a Conventional Terrestrial Reference System
(CTRS) as a reference system resulting from a [time-dependent] spatial
rotation of the geocentric reference system defined by the IAU 1991
Recommendations, the spatial rotation being chosen  such  that the CTRS
has no global residual rotation with respect to horizontal motions at
the Earth's surface.  The coordinate time of the CTRS coincides with
$TCG$.

IAU Resolution C7 (1994) recommends that the epoch J2000 as well as the
Julian (ephemeris) day are defined in $TT$.

IAU Resolution B6 (1997) has supplemented the framework by one more
recommendation stating that no scaling of spatial axes should be
applied in any reference system (even if a scaled time coordinate like
$TT$ is used). Note, however, that this Resolution has been ignored in
the construction of the International Terrestrial Reference Frame
(ITRF) which is defined not with the GCRS spatial coordinates $\vX$ but
with scaled coordinates $\vX_{TT} = (1 - L_G)\, \vX$.

\section{The IAU 2000 Resolutions on relativity}
\label{Section-IAU-2000}

The IAU 1991 framework is unsatisfactory from many points of view. The
Einstein-Infeld-Hoffmann equations of motion which have been used since
the 70s to construct the JPL numerical ephemerides of planetary motion
cannot be derived from the metric (\ref{iau_metric}). In other words
for the motion of massive solar system bodies metric (\ref{iau_metric})
is not the post-Newtonian metric of Einstein's theory of gravity. In
the years prior to the XXIII General Assembly  in Kyoto (1997) it
became obvious that the IAU 1991 set of recommendations concerning relativity
in astrometry, celestial mechanics and metrology is not sufficient for
achievable accuracies. Especially with respect to planned astrometric
missions with microarcsecond accuracies extended and improved resolutions
became indispensable. For that reason the IAU Working Group ``Relativity
for astrometry and celestial mechanics'' together with the BIPM/IAU
Joint Committee on relativity for space-time reference systems and
metrology suggested such an extended set of Resolutions (B1.3 - B1.5
and B1.9) that was finally adopted at the IAU General Assembly in
Manchester in the year 2000. The relevant Resolutions can be found in
Appendix A. It is clear that because of their brevity they need
additional explanations and one has to show how they work in practice.
This paper now presents a detailed explanatory supplement for these IAU
2000 Resolutions.

\subsection{The role of the two Celestial Reference Systems, BCRS and
GCRS}
\label{Section-two-CRS}

Some of the reasons why two different celestial astronomical reference
systems have to be introduced have already been mentioned in the
introduction. Here we would like to deepen this discussion in several
respects. It is clear that for many applications in the fields of
astrometry, celestial mechanics, geodynamics, geodesy etc.\ some
quasi-inertial or 'space-fixed' reference system has to be introduced.
Resolution B1.3 defines actually {\it two} different Celestial
Reference Systems: the Barycentric Celestial Reference System (BCRS)
and the Geocentric Celestial Reference System (GCRS).

In {\it Newtonian theory}  one can easily introduce inertial space-time
coordinates that cover the entire universe. Such inertial coordinates
in Newton's theory are unique up to the choice of origin, scales, the
orientation of spatial axes and up to a constant velocity of origin. In
astronomy {\it conceptually} we may talk about two different relevant
celestial systems: a barycentric  and a geocentric one that basically
serve different purposes. The barycentric celestial system is
considered to be inertial (external galactic and extragalactic
matter being normally neglected)
and is used for solar system ephemerides, for
concepts such as an ecliptic, for interplanetary spacecraft navigation
etc. The positions of remote objects can be defined in that system. The
barycentric celestial system presents the fundamental astrometric
system where concepts as 'proper motion' or 'radial velocity' can be
defined.

On the other side the geocentric celestial system might be called
quasi-inertial since  the spatial axes are non-rotating in the
Newtonian absolute sense whereas the geocenter is accelerated. It is
employed for the description of physical processes in the vicinity of
the Earth, for satellite theory, the dynamics of the Earth including
Earth's rotation, etc. It is also used for the introduction of concepts
like equator and the ITRS. Let us denote the time and space coordinates
of the barycentric celestial system by $(t,\vx)$, those of the
geocentric celestial system by $(T,\vX)$. In Newton's framework the
relation between these two sets of coordinates is trivial

$$
T = t, \qquad \vX = \vx - \vx_E(t) \, ,
$$

\noindent
where $\vx_E(t)$ denotes the barycentric position of the geocenter.
Because these relations are so trivial for some purposes the
barycentric and the geocentric celestial systems are not always clearly
distinguished in the Newtonian framework.

Of course for astrometric problems one always distinguished between the
two celestial systems and apparent places of stars from true
(barycentric)  places. However, annual parallax and aberration were
merely understood as correction terms that have to be applied to get
the 'true' positions for the realization of the astronomical quasi-inertial,
space-fixed celestial system. Note that the definition of the classical
astronomical $(\alpha,\delta)$ system uses concepts from both systems:
some ecliptic from the barycentric celestial system and some Earth's
rotation pole, Celestial Ephemeris Pole (CEP) or
Celestial Intermediate Pole (CIP), and its corresponding equator from the
geocentric celestial system.

In {\it Relativity theory} the situation is actually more complicated. Even in
the absence of gravitational fields and a uniformly moving geocenter
the two coordinate systems are related by a 4-dimensional
Lorentz transformation of Special Relativity. In our solar system BCRS
and GCRS coordinates are related by a complicated 4-dimensional
space-time transformation (a generalized Lorentz transformation) that
also contains acceleration terms and gravitational potentials. This
implies that the two astronomical Reference Systems, the BCRS and the
GCRS are actually quite different. This has profound consequences for a
lot of classical astronomical concepts.

The BCRS is the basic astrometric celestial reference system. Usually
one considers the solar system to be isolated, i.e., one ignores all
matter and fields outside the system and assumes the gravitational
potentials to vanish far from the system. It is obvious that ignoring
the Galaxy and extragalactic objects is an unphysical idealization for
several specific questions that, however, will not be touched here. If
the solar system is considered to be isolated we might follow
light-rays from some very remote source back in time to the region
$\vert\vx\vert \rightarrow \infty$ that might be called the celestial
sphere. In the vicinity of the celestial sphere a certain light ray
defines spherical angles that might appear as catalog values.
Actually for reference stars the physical distance from the Earth
usually  plays a role. In that case we might associate any star with a
corresponding BCRS coordinate position $\vx_*$ that will be a function
of $TCB$. From this position vector spherical angles
$(\alpha_*,\delta_*)$ can be introduced in a very simple manner by

\begin{equation}
{\vx_* \over \vert \vx_* \vert} = \left(
\begin{array}{c}
\cos\alpha_* \cos\delta_* \\
\sin\alpha_* \cos\delta_*  \\
\sin \delta_*
\end{array}
\right)
\end{equation}

\noindent
that can be considered as catalog values. If the coordinate distance
of some source tends to infinity the two constructions for an
astrometric position will coincide. From $\vx_*(t)$ quantities such as
'proper motion' or 'radial velocity' can be defined as coordinate
quantities in the BCRS. Note that the problem of 'radial velocity' has
exhaustively been discussed by \citet{Lindegren:Dravins:2003}
(see also IAU 2000 Resolutions C1 and C2 \citep{Rickman:2001}).
Other fields of application of the BCRS are
solar system ephemerides, interplanetary navigation, etc.

The definitions of the BCRS given by the IAU 2000 Resolution B1.3 do
not fix the orientation of spatial axes uniquely but only up to some
constant, time independent rotation matrix about the origin. One
natural choice of orientation is provided by the International
Celestial Reference System (ICRS). Actually for the construction of the
International Celestial Reference Frame (ICRF) and its optical
counterpart, the Hipparcos catalog, the recommended form of the
barycentric metric tensor has already been used explicitly in the
underlying models. This implies that a set of definitions that fix the
ICRS completely contains the BCRS definitions.

There might be other useful possibilities for the orientation of
barycentric spatial coordinates. One possibility is  the orientation
according to some ecliptic ${\cal E}_0$ at a certain epoch $t_0$
defined by corresponding solar system ephemerides. Such an ecliptic
would coincide with the $x$-$y$ plane of a BCRS[${\cal E}_0$] that
might be useful for reasons of historical continuity.

On the other hand quantities and concepts related with the physics in
the immediate vicinity of the Earth should be formulated in the GCRS.
This concerns the gravity field of the Earth itself, satellite theory
and especially applies to theories of Earth's rotation and their
parameters. Clearly the spatial GCRS coordinates $\vX$ can be used to
define corresponding unit vectors at the geocenter that might be
employed  to compute spherical angles $(\alpha_{\rm GCRS},\delta_{\rm
GCRS})$ which might be called 'geocentric places'. Note, however, that
the coordinates of the remote astronomical sources are defined in the
BCRS only. The calculated GCRS places $(\alpha_{\rm GCRS},\delta_{\rm
GCRS})$ are determined by incident light rays at the geocenter. They
differ from corresponding ICRS $(\alpha,\delta)$ values because of
annual aberration, annual parallax and gravitational light deflection
due to the gravitational fields of the solar system bodies (apart from
the Earth) and are independent of Earth's rotation. If these GCRS
places, however, will ever play a role in practice is not clear.

In the past apparent places of stars that were annually published e.g.,
in the 'Apparent Places of Fundamental Stars' played a role for certain
problems. These places are related with the old traditional
astronomical reference system, i.e., with some equator and equinox of
date. Now with the ICRS we have a highly precise  astronomical
reference system that is basically independent of Earth's rotation
parameters and their determination. For several applications, however,
the introduction of quantities like apparent places might still be
useful especially if there is a reference to the local plumb line,
i.e., to the zenith and the astronomical (or nautical) triangle can be
employed. In that case Resolutions B1.7 and B1.8
\citep{IAU:2001,Rickman:2001} come into play. These two Resolutions define some
intermediate system that can be used for the definition of an
intermediate position $(\alpha_{\rm inter}, \delta_{\rm inter})$ by the
Celestial Intermediate Pole (CIP) and the Celestial Ephemeris Origin
(CEO). Such intermediate position can be considered as modern version
of the apparent place, defined in the GCRS.

For astrometry at microarcsecond accuracies neither GCRS places nor
intermediate places likely will play a role. To avoid problems related
with non-linearities it is  simpler to use an overall BCRS picture to
describe not only the light-rays and the motion of gravitating bodies
but also the trajectory of an observer. In that case only catalog and
observed positions will be of importance.

\subsection{The Barycentric Celestial Reference System}
\label{Section-BCRS}

Resolution B1.3 concerns the definition of Barycentric Celestial
Reference System (BCRS) and Geocentric Celestial Reference System
(GCRS). The BCRS is defined with coordinates $(ct,x^i) = x^\mu$ where
$t =  TCB$. The BCRS is a particular version of the barycentric
reference system of the solar system. The Resolution recommends to
write the metric tensor of the BCRS in the form

\begin{eqnarray}
\label{BCRS_metric}
g_{00} &=& -1 + {2w \over c^2} - {2w^2 \over c^4} + \OO5, \nonumber\\
g_{0i} &=& -{4 \over c^3} w^i+\OO5, \label{bary_metric}\\
g_{ij} &=& \delta_{ij}\left(1 + {2 \over c^2}w \right) + \OO4 \, . \nonumber
\end{eqnarray}

A comparison reveals that this form of the metric presents an extension
of (2). Whereas the old form contains only, the Newtonian potential $U$
the new one contains a scalar potential $w$ and a vector potential
$w^i$.

Actually the equations for $g_{00}$ and $g_{0i}$ from
(\ref{BCRS_metric}) without the order symbols $\OO5$ are always correct
and can be simply considered as definitions of $w$ and $w^i$ in terms
of $g_{00}$ and $g_{0i}$. In contrast to the concrete form of the
resolution we have added order symbols in (\ref{BCRS_metric}). E.g.,
for $g_{00}$ the order symbol indicates that terms of order $c^{-5}$
will systematically be neglected as stated in the notes to the
Resolution. With these forms for $g_{00}$ and $g_{0i}$ one finds that
spatially isotropic coordinates $x^i$ exist such that  $g_{ij}$ from
equation (\ref{BCRS_metric}) with the potential $w$ from $g_{00}$
solves Einstein's field equations to first post-Newtonian order. Note
that the form of (\ref{bary_metric}) implies the barycentric spatial
coordinates $x^i$ satisfy the harmonic gauge condition \citep[see
e.g.][]{Brumberg:Kopeikin:1989,Damour:Soffel:Xu:1991}. At this point,
because of the freedom in the  time coordinate, many different 'time
gauge conditions' are still possible. The Resolution proceeds by
recommending a specific kind of  space and time {\it harmonic gauge}.
One argument in favor of the harmonic gauge is that tremendous work on
General Relativity has been done with the harmonic gauge that was found
to be a useful and simplifying gauge for many kinds of applications.
Moreover, the harmonic gauge condition
\citep[e.g.,][]{Weinberg:1972,Fock:1959}

\begin{equation} \label{harmonic}
g^{\mu\nu} \Gamma^\lambda_{\mu\nu} = 0, \,
\end{equation}

\noindent
where $\Gamma^\lambda_{\mu\nu}$ are the Christoffel-symbols of the
metric tensor, is not restricted to some post-Newtonian approximation,
but can be defined in Einstein's theory of gravity without any
approximations. This may be important for future refinements of the IAU
framework. With the harmonic gauge condition the post-Newtonian
Einstein field equations take the form

\begin{eqnarray}\label{PN_fielda}
\left( - {1 \over c^2} {\partial^2 \over \partial t^2}
+ \Delta \right) w &=& - 4 \pi G \sigma + \OO4,
\\
\label{PN_fieldb}
\Delta w^i &=& - 4 \pi G \sigma^i + \OO2 \, .
\end{eqnarray}

\noindent
Here $\sigma$ and $\sigma^i$ are the gravitational mass and
mass current density, respectively. Mathematically they are related to
the {\it energy-momentum tensor} $T^{\mu\nu}$ by

\begin{equation} \label{sigma}
\sigma = {1\over c^2}\left(T^{00} + T^{ss}\right), \quad
\sigma^i = {1\over c}\ T^{0i}\, .
\end{equation}

\noindent
The energy-momentum tensor $T^{\mu\nu}$ generalizes the density $\rho$
of the Poisson equation (\ref{poisson}). In relativity, energy density,
pressure and stresses all act as source of the gravitational field.
This implies that different  kinds of energy contribute to the
gravitational sources: kinetic energy, gravitational potential energy,
energy of deformation  etc., but since the kinetic energy depends upon
the state of motion of the matter, the energy-momentum tensor that
really acts as gravitational source shows a nontrivial transformation
behaviour if we go from one reference system to another. In  practice,
however, the energy-momentum tensor will usually not appear explicitly.
This is because the gravitational potentials $w$ and $w^i$ from
(\ref{PN_fielda})--(\ref{PN_fieldb}) are completely determined by
$\sigma$ and $\sigma^i$ which can be considered as primary quantities.
If we deal with problems where gravitational fields play a role only
{\it outside} of astronomical bodies  and admit a useful convergent
expansion in terms of multipole moments (potential coefficients) only
corresponding integral characteristics of the bodies like masses,
quadrupole moments etc. show up explicitly, which are defined in terms
of $\sigma$ and $\sigma^i$  and whose numerical values will be fixed by
observations. Because of (\ref{PN_fieldb}) $w^i$ is sometimes called
the gravitomagnetic potential since it results from mass currents
(moving or rotating masses) just as the electromagnetic vector
potential
results from electric currents in Maxwell's theory of electromagnetism.

Equation  (\ref{PN_fielda}) generalizes the Poisson equation
(\ref{poisson}), hence the scalar potential $w$ presents a relativistic
generalization of the Newtonian potential $U$. Because of problems
related with  homogeneous solutions and boundary conditions
mathematically it is clear that these differential equations do not fix
the harmonic solutions uniquely. Assuming space-time to be
asymptotically flat (no gravitational fields far from the system),
i.e.,

\begin{equation}
\lim_{{r \rightarrow \infty}\atop t={\rm const}} g_{\mu\nu}
= {\rm diag}(-1,+1,+1,+1)
\end{equation}

\noindent
the recommended solution reads

\begin{eqnarray}\label{fielda}
w(t,\ve{x}) &=& G \int d^3 x' \,
{\sigma(t, \ve{x}') \over \vert \ve{x} - \ve{x}' \vert}
 + {1 \over 2c^2} G {\partial^2 \over \partial t^2}
\int d^3 x' \, \sigma(t,\ve{x}') \vert \ve{x} - \ve{x}' \vert \, ,
\\
\label{fieldb}
w^i(t,\ve{x}) &=& G \int d^3 x' {\sigma^i (t,\ve{x}') \over
\vert\ve{x} - \ve{x}' \vert } \, .
\end{eqnarray}

\noindent
It is obvious that the second time derivative term in (\ref{fielda})
results from the corresponding operator in the field equation
(\ref{PN_fielda}). This operator modifies the Laplacian from the
Newtonian Poisson equation to the d'Alembertian and the similarity
between the harmonic post-Newtonian field equations and Maxwell's equations of
electromagnetism in the Lorentz gauge becomes obvious (actually one
might replace the Laplacian by the d'Alembertian in (\ref{PN_fieldb})
to post-Newtonian accuracy). From Maxwell's theory it is well known
that the retarded potential solves the corresponding field equation

\begin{equation}
\label{ret}
w_{\rm ret} (t,x^i) = G \int d^3 x' \,
{\sigma(t_{\rm ret}, \ve{x}') \over \vert \ve{x} - \ve{x}' \vert}
\end{equation}
with
\begin{equation}
t_{\rm ret} = t - {\vert \vx - \vx' \vert \over c} \, .
\end{equation}

\noindent
One might then expand the retarded potential in terms of $1/c$. Note,
that such an expansion also yields a term proportional to $1/c$. If we
stay within the first post-Newtonian approximation these $1/c$ terms
vanish due to the Newtonian mass conservation law. Such odd powers of
$1/c$ indicate time asymmetric terms, i.e., they break time reversal
symmetry. It is well known that such time asymmetric terms appear only
to higher  post-Newtonian order and will not be considered here.
For that reason the retarded potential (\ref{ret}) leads to the
recommended solution above.

Comparing the form of the metric tensor in
(\ref{BCRS_metric}) with other forms that can be found in the
literature (e.g., Will 1993) one might get the erroneous impression
that something is missing in (\ref{BCRS_metric}), which is
not the case. If matter is discribed by some fluid model then formally
($w, w^i$) might be split into various pieces resulting from kinetic
energy, gravitational potential energy, specific internal energy
density, pressure etc.\ and the equivalence of our form of the metric
tensor e.g.\ with that given in Will (1993) can be shown. This is
explicitly demonstrated in Appendix B.

The point, however, is that a split of  $(\sigma,\sigma^i)$ of our
metric potentials $(w,w^i)$ or of the metric tensor itself into various
pieces is usually unnecessary. If only gravitational fields outside the
relevant bodies play a role (as is typically the case in celestial
mechanics and astrometry) it is advantageous  to keep such  pieces
together, since it will be the sum that determines the observables. One
might argue, $U$ is the 'Newtonian potential' and the rest can be
identified as 'relativistic corrections'. This way of thinking,
however, can be very misleading and presents a source of errors. As has
been shown in the literature
\citep[e.g.,][]{Damour:Soffel:Xu:1991,Damour:Soffel:Xu:1993} suitably
defined potential coefficients based upon $w$ (not $U$) and $w^i$ can
be introduced that can be determined from satellite data. From a more
theoretical point of view the introduction of $(w,w^i)$ has the
advantage that the field equations (\ref{PN_fielda})--(\ref{PN_fieldb})
are  formally {\it linear}, although the corresponding metric is not
(because of the $w^2$-term). We used the word 'formally' since $\sigma$
depends upon $w$ implicitly. This nonlinearity has been explicitly
treated, e.g., by \citet{Brumberg:Kopeikin:1989}, but this dependence
becomes practically irrelevant if the fields outside of some matter
distribution are parametrized by means of potential coefficients. This
linearity implies that for an ensemble of $N$-bodies

\begin{equation}
\label{wsum}
w(t,\vx) = \sum_{A=1}^N w_A(t,\vx) \, , \qquad
w^i(t,\vx) = \sum_{A=1}^N w_A^i(t,\vx)\, ,
\end{equation}

\noindent
where the index $A$ indicates the contribution related with body $A$
where the integrals have to be taken over the support of body $A$ only.
This linearity, however, does not imply that body-body interaction
terms have been neglected. If written explicitly $w_A$ will in general
contain contributions from bodies $B \not= A$ (see e.g., Eq.
(\ref{delta-A})).

The BCRS metric tensor from the IAU 2000 Resolution B1.3 extends the form of
the metric tensor given in the IAU 1991 Resolutions such that its
accuracy is sufficient for most applications in the next years. Note
that an extension of the old metric (\ref{iau_metric}) is necessary
(and has been in use for decades) for the derivation of the
relativistic  equations that form the basis of any modern solar system
ephemeris (such as the JPL DE ephemerides). Resolution B1.3 formalises
this extension.

\subsection{The Geocentic Celestial Reference System}
\label{Section-GCRS}

Resolution B1.3 continues to define the GCRS which represents a
particular version of the local geocentric reference system for the
Earth. Its spatial coordinates $X^a$ are kinematically non-rotating
with respect to the barycentric ones \citep[see
e.g.,][]{Brumberg:Kopeikin:1989,Klioner:Soffel:1998}. The geocentric
coordinates are denoted by $(T,\vX)$, where $T = TCG$.
In the relation between $x^i$ and $X^a$
from Resolution B1.3 let us replace the unit matrix $\delta_{ai}$ by a general rotation
matrix $R_{ai}$

\begin{displaymath}
X^a = R_{ai}
\left[ r_E^i + {1 \over c^2}
\left(  \dots \right) \right]
+ \cO(c^{-4}) \, ,
\end{displaymath}

\noindent
where $\vr_E = \vx - \vx_E$. If the two sets of spatial coordinates are
aligned for all times, i.e., if $R_{ai} = \delta_{ai}$ as is the case
for the GCRS spatial coordinates, then $X^a$ is defined to be
kinematically non-rotating with respect to the barycentric spatial
coordinates $x^i$. The Resolution recommends writing the metric tensor
of the GCRS in the same form as the barycentric one but with potentials
$W(T,\vX)$ and $W^a(T,\vX)$. Explicitly,

\begin{eqnarray}\label{geo_metric}
G_{00} &=& -1 + {2W \over c^2} - {2W^2 \over c^4} + \OO5, \nonumber\\
G_{0a} &=& -{4 \over c^3} W^a+\OO5, \\
G_{ab} &=& \delta_{ab}\left(1 + {2 \over c^2}W \right) + \OO4 \nonumber
\end{eqnarray}

\noindent
and the geocentric field equations formally look the same as the
barycentric ones (\ref{PN_fielda})--(\ref{PN_fieldb}) but with all
variables referred to the GCRS. Again one decisive advantage of this
recommendation is the formal linearity of the field equations. This
linearity admits a unique split of the geocentric metric into a part
coming from the Earth itself and a remaining part resulting from
inertial and tidal forces. Therefore it is recommended to split the
potentials $W$ and $W^a$ according to

\begin{equation}
W(T,{\bf X}) = W_E(T,{\bf X}) + W_{\rm ext}(T,{\bf X}), \qquad
W^a(T,{\bf X}) = W_E^a(T,{\bf X}) + W_{\rm ext}^a(T,{\bf X}).
\end{equation}

\noindent
The Earth's potentials $W_E$ and $W_E^a$ are defined in the same way as
$w_E$ and $w_E^a$, (i.e., equations (\ref{fielda})--(\ref{fieldb}) with
integrals taken over the volume of the whole Earth) but with quantities
calculated in the GCRS. Outside the Earth the potentials $(W,W^a)$
admit a power series expansion in terms of $R \equiv \vert\vX\vert$ and
all negative powers of $R$ are contained in $W_E$ and $W_E^a$. For that
reason the Earth's potentials admit  multipole expansions that look
very similar to the Newtonian ones. This point will be discussed below
in more detail.

It is useful to split the external potentials $W_{\rm ext}$ and $W_{\rm
ext}^a$ further. They can be written in the form

\begin{equation}
W_{\rm ext} = W_{\rm tidal} + W_{\rm iner}, \qquad
W_{\rm ext}^a = W_{\rm tidal}^a + W_{\rm iner}^a \, ,
\end{equation}

\noindent
where the tidal terms are at least quadratic in $X^a$ and the inertial
contributions $W_{\rm iner}$ and $W_{\rm iner}^a$ are just  linear  in
$X^a$. Explicitly,

\begin{eqnarray}\label{W-iner}
W_{\rm iner}&=& Q_a\,X^a,
\nonumber\\
\label{W-iner-a}
W^a_{\rm iner} &=& - {1\over 4}\,c^2
\,\epsilon_{abc} \Omega^b_{\rm iner} \,X^c
\, .
\end{eqnarray}

\noindent
Mathematically the $Q_a$ term is related with the 4-acceleration of the
geocenter in the external gravitational field, a quantity that vanishes
for a purely spherical and non-rotating Earth (for a mass monopole more
precisely) that moves along a geodesic in the external gravitational
field. The $Q_a$ term therefore results from the coupling of higher
order multipole moments of the Earth to the external tidal
gravitational fields (to the external curvature tensor of space-time
in mathematical language). $Q_a$ characterizes the deviation of the actual
worldline of the origin of the GCRS from a geodesic in the external
gravitational field. With

\begin{displaymath}
w_{\rm ext}(t,\vx) = \sum_{A\not= E} w_A(t,\vx) \, ,
\qquad
w^i_{\rm ext}(t,\vx) = \sum_{A\not= E} w^i_A(t,\vx) \,
\end{displaymath}

\noindent
to Newtonian order $Q_a$ is given by

\begin{equation}\label{}
Q_a = \delta_{ai}\,\left({\partial\over\partial x^i}\,
w_{\rm ext}(\vx_E) - a_E^i\right)\, .
\end{equation}

\noindent
Here, $x_E^i(t), v_E^i(t) = dx_E^i/dt$ and $a_E^i={dv^i_E/dt}$ are the
barycentric coordinate position, velocity and acceleration of the
origin of the GCRS (geocenter). The appearance of $\delta_{ai} $
results from the fact that the GCRS is defined as
kinematically non-rotating with respect to the BCRS. The reason to
retain $\delta_{ai}$ in the transformations here and below results from
the desire to distinguish between BCRS (spatial indices taken from the
second part of the latin alphabet, starting with the letter $i$) and
GCRS quantities (spatial indices taken from the first part of the
alphabet).

The full post-Newtonian expression for $Q_a$ (denoted by $G_a(T)$ in
the Damour-Soffel-Xu papers) can be derived from (6.30a) of
\citet{Damour:Soffel:Xu:1991}. To get an idea about orders of magnitude
the absolute value of $Q_a$ due to the action of the Moon is of order
$4 \times 10^{-11}\,$m/s$^2$ \citep{Kopeikin:1991}.

The term $W^a_{\rm iner}$ describes a relativistic Coriolis force due
to the rotation of the GCRS with respect to a dynamically non-rotating
geocentric reference system. Such a rotation has several components,
often referred to as geodesic, Lense-Thirring and Thomas
precessions

\begin{equation}\label{Omega-iner}
\vOmega_{\rm iner} =
\vOmega_{\rm GP}
+ \vOmega_{\rm LTP}
+ \vOmega_{\rm TP}
\end{equation}

\noindent
with

\begin{eqnarray}
\label{Omega-GP}
\vOmega_{\rm GP} &=& -{3\over 2 c^2}\, \vv_E \times
\nabla
w_{\rm ext}(\ve{x}_E),
\nonumber\\
&& \nonumber\\
\label{Omega-LT}
\vOmega_{\rm LTP} &=& -{2\over c^2}\, \nabla \times \vw_{\rm ext}(\ve{x}_E),
\\
&& \nonumber\\
\label{Omega-TP}
\vOmega_{\rm TP} &=& -{1\over 2c^2}\, \vv_E \times \vQ \, ,
\nonumber
\end{eqnarray}

\noindent
in obvious notation. As a relativistic precession, the geodesic
precession, $\vOmega_{\rm GP}$, is proportional to $1/c^2$. It is also
proportional to the barycentric coordinate velocity $v_E$ and the
gradient of the external gravitational scalar potential $w_{\rm ext}$
at the geocenter (the barycentric coordinate acceleration of the
geocenter to sufficient accuracy). The order of magnitude is given by
$\vert\vOmega_{\rm GP} \vert$$ \sim 1.5 (v_E/c)(G M_S/c^2 AU)(c/AU)
\sim 3 \times 10^{-15}\,$s$^{-1} \sim 2 \arcsec$/century. Thomas
precession is also proportional to $1/c^2$ and the barycentric
coordinate velocity of the geocenter but also to the geodesic deviation
term $Q_a$. The order of magnitude of Thomas precession is $\vert
\vOmega_{\rm TP} \vert \sim 0.5 (v_E/c) \vert \vQ \vert/c \sim 7 \times
10^{-24} \, {\rm s}^{-1} \sim 4 \times 10^{-9}{}\arcsec$/century, i.e.,
negligible with respect to geodesic precession.

Finally the Lense-Thirring precession results from the gradient of the
external gravitomagnetic potential at the geocenter. If we consider
some spherically symmetric solar system body $A$, then the
gravitomagnetic potential $W^a_A$ of it is given by (see
(\ref{Wa-spin-only}) below)

\begin{displaymath}
W^a_A = {G \over 2}
{(\vS_A \times \vX)^a \over R^3}
\end{displaymath}

\noindent
in its own local rest frame. Transformation into the BCRS according to
the rule indicated below in (\ref{wEi-WEa}) leads to

\begin{displaymath}
w_A^i(t,\vx) = G \left[
{(\vS_A \times {\bf r}_A)^i \over 2 r_A^3} + {M_A \over r_A} v_A^i
\right] \, ,
\end{displaymath}

\noindent
where ${\bf r}_A \equiv \vx - \vx_A$ and $\vv_A$ is the barycentric
velocity of body $A$. In our case the spin and motion of our Sun and
Moon will give the dominant contributions to $\vOmega_{\rm LTP}$:
$\vert \vOmega_{\rm LTP} \vert \sim 2 \times 10^{-3}{}\arcsec$/century.

The definition of the GCRS implies that the spatial GCRS coordinates
$\vX$ are kinematically non-rotating with respect to the BCRS ones,
$\vx$ (as indicated by the $\delta_{ai}$-term in Resolution B1.3).
Because of geodesic precession locally inertial coordinates precess
with respect to  the GCRS by an amount of $\vert\vOmega_{\rm iner}
\vert = 1.9198\arcsec/$century \citep{Brumberg:et:al:1992}. Let us
forget about the mass of the Earth and imagine a torque free gyroscope
at the geocenter, moving along the actual trajectory of the geocenter.
It will precess by this amount in our GCRS. Since the GCRS does not
present a locally inertial reference system Coriolis forces caused by
geodesic precession-nutation appear in every GCRS dynamical equation of
motion, e.g., of Earth's satellites. As recommended by
\citet{IERS:2003} these additional forces should be taken into account.
Moreover, geodesic precession-nutation has to be considered in the
precession-nutation model formulated in the GCRS. E.g., the basic
post-Newtonian equation of Earth's intrinsic angular momentum $\vS$
reads

\begin{equation}
{d\vS \over dT} + \vOmega_{\rm iner} \times \vS = \vD \, ,
\end{equation}

\noindent
where $\vD$ is the external torque \citep{Damour:Soffel:Xu:1993}. As
long as observations of Earth's orientation parameters are referred to
the GCRS they will contain geodesic precession-nutation automatically.

Because of the eccentricity of the Earth orbit the leading term in
$\vOmega_{\rm GP}$ has an annual and a semi-annual part that leads to
geodesic nutation in longitude with

\begin{equation}
\Delta \psi_{\rm GP} = 0.153 \sin l' + 0.002 \sin 2 l' \, ,
\end{equation}

\noindent
where the amplitudes are in mas and $l'$ is the mean anomaly of the
Earth-Moon barycenter
\citep{Fukushima:1991,Brumberg:et:al:1992,Bois:Vokrouhlicky:1995}.

$W_{\rm tidal}$ is a generalization of the Newtonian tidal potential

\begin{equation}\label{W-tidal}
W^{\rm Newton}_{\rm tidal}(T,\vX) =
w_{\rm ext}(\ve{x}_E + \vX) - w_{\rm ext}(\ve{x}_E) - \vX \cdot
\nabla  w_{\rm ext}(\ve{x}_E)  \, .
\end{equation}

\noindent
Full post-Newtonian expressions for $W_{\rm tidal}$ and $W^a_{\rm
tidal}$ can be found in \citet{Damour:Soffel:Xu:1992}. There $W_{\rm
ext}$ is denoted by $\overline W$ and a tidal expansion in powers of
local spatial coordinates by means of suitably defined tidal moments is
given in (4.15) of \citet{Damour:Soffel:Xu:1992}. Expressions for
$W_{\rm tidal}$ and $W^a_{\rm tidal}$ in closed form are given in
\citep{Klioner:Voinov:1993}. The quadratic term as dominant term of
$W_{\rm tidal}$ reads

\begin{equation}
\biggl.W_{\rm tidal}\biggr|_{l=2} = {1 \over 2} G^{\rm tidal}_{ab}\, X^a X^b \, .
\end{equation}

\noindent
If the external bodies are taken as mass monopoles the explicit
expression for $G^{\rm tidal}_{ab}$ (not to be confused
with the GCRS metric tensor) is given in (3.23) of
\citet{Damour:Soffel:Xu:1994}. Higher-order terms in this approximation
can be found in \citet{Klioner:Soffel:Xu:Wu:2000}.

Finally, the local gravitational potentials $W_E$ and $W_E^a$ of the
Earth are related to the barycentric gravitational potentials $w_E$ and
$w^i_E$ by $(\delta_i^a = \delta_a^i = \delta_{ai})$

\begin{eqnarray}\label{WE-wE}
W_E(T,\ve{X})&=&w_E(t,\ve{x})\,\left(1+{2\over c^2}\, v_E^2\right)-
{4\over c^2}\,v_E^i\,w^i_E(t,\ve{x})+\OO4,
\nonumber\\\label{WEa-wEi}
W^a_E(T,\ve{X})&=&\delta^a_i\,\left(w^i_E(t,\ve{x})
-v_E^i\,w_E(t,\ve{x})\right) + \OO2
\end{eqnarray}

\noindent
or by the inverse transformation

\begin{eqnarray}\label{wE-WE}
w_E(t,\ve{x})&=&W_E(T,\ve{X})\,\left(1+{2\over c^2}\, v_E^2\right) +
{4\over c^2}\,\delta_{ia} v_E^i\, W^a_E(T,\ve{X})+\OO4,
\nonumber\\
\label{wEi-WEa}
w^i_E(t,\ve{x}) &=& \delta^i_a\, W^a_E(T,\ve{X})
 + v_E^i\,W_E(T,\ve{X}) +\OO2.
\end{eqnarray}

\noindent
The relations between the geocentric gravitational potentials $W$ and
$W^a$, and the barycentric ones $w$ and $w^i$ follow from the
coordinate transformations between the BCRS and GCRS discussed below.

\subsection{Coordinate transformations}
\label{Section-trans}

The metric tensors in the BCRS and GCRS allow one to derive the rules
for the transformations between the BCRS coordinates $x^\mu$
and the GCRS ones $X^\alpha$ from the tensorial  transformation rules.
It is obvious that these
transformations can be written in two equivalent forms: i) as
$x^\mu(T,X^a)$ or ii) as $X^\alpha(t,x^i)$. Whereas the first form was
used in the Damour-Soffel-Xu formalism \citep{Damour:Soffel:Xu:1991,
Damour:Soffel:Xu:1992,Damour:Soffel:Xu:1993,Damour:Soffel:Xu:1994}, the
second one was presented in the Brumberg-Kopeikin formalism
\citep{Brumberg:Kopeikin:1989,Kopeikin:1988,Brumberg:1991,
Klioner:Voinov:1993}. It should be pointed out that the transformation
from one version to the other is not so trivial because of the
barycentric coordinate position of the geocenter that appears in the
first form as function of $TCG$ and as function of $TCB$ in the second
one. In Resolution B1.3 $T = TCG$ and $X^a $ are presented as functions
of $t = TCB$ and $x^i$. The explicit form of the transformations is
given in the text of Resolution B1.3 (see, Appendix A below). Apart
from the terms of order ${\cal O}(|\ve{X}|^3)$ that appear in the time
transformation of order $\OO4$ all terms can be obtained from the
results derived by \citet{Kopeikin:1988} and
\citet{Damour:Soffel:Xu:1991}. The cubic and higher order terms  in $|\ve{X}|$
as represented by the function $C$ have been
derived by \citet{Kopeikin:1988} and analyzed in full detail by
\citet{Klioner:Voinov:1993}. As is also clear from
\citet{Klioner:Voinov:1993} the expression for $C$ is not unique, but
only constrained by the gauge and field equations so that the simplest
possibility is an expression for $C$ containing cubic terms only. It is
this simplest expression that is recommended in Resolution B1.3.

The full 4-dimensional coordinate transformation is just an extension
of the usual Lorentz transformation. Indeed, if we neglect all
gravitational fields and acceleration terms then the coordinate
transformation in Resolution B1.3 can be written in the form ($\vr=
\vx - \vx_E(t)$, $\beta = v/c = {\rm const.}$)

\begin{eqnarray}
T &=& t \left( 1 - {1 \over 2} \beta^2 - {1 \over 8} \beta^4 \right)
- \left( 1 + {1 \over 2} \beta^2 \right) {\vv \cdot \vr \over c^2} +
\OO6, \nonumber\\
\vX &=& \vr + {1 \over 2}
( \vv \cdot \vr) {\vv \over c^2} + \OO4 \, .
\end{eqnarray}

\noindent
If we now write $\vx_E(t) = \vv t$ we obtain

\begin{eqnarray}
T &=& t \left( 1 + {1 \over 2} \beta^2 + {3 \over 8} \beta^4 \right)
- \left( 1 + {1 \over 2} \beta^2 \right) {\vv \cdot \vx \over c^2} +
\OO6, \nonumber \\
\vX &=& \vx - \left( 1 + {1 \over 2} \beta^2 \right) \vv t
 + {1 \over 2}
( \vv \cdot \vx) {\vv \over c^2} + \OO4 \, .
\end{eqnarray}

\noindent
which is nothing but a Lorentz tranformation from Special Relativity
Theory

\begin{displaymath}
T = \gamma \,\left(t - {\vv \cdot \vx \over c^2}\right) \, , \qquad
\vX = \vx - \gamma \vv t + {(\gamma - 1) \over v^2} (\vv \cdot \vx)
\vv
\end{displaymath}

\noindent
in the corresponding  approximation since

\begin{displaymath}
\gamma \equiv (1 - \beta^2)^{-1/2} = 1 + {1 \over 2}
\beta^2 + {3 \over 8} \beta^4 + \OO6 \, .
\end{displaymath}

\noindent
Note that the inverse transformations are obtained simply by replacing
$(t,\vx)$ by $(T,\vX)$ and the velocity $\vv$ by $- \vv$.

Neglecting the $1/c^4$ terms in the $T - t$ relation given in
Resolution B1.3 one gets

\begin{equation}
T = t - {1 \over c^2}
\left(\int^t_{t_0}\left({v_E^2 \over 2} + w_{\rm ext}(\vx_E)\right)dt
+ v_E^i r_E^i
\right) + \OO4
\end{equation}

\noindent
which reduces to the old recommendation  (\ref{oldtcb-tcg}) since $t =
TCB$, $T = TCG$ and $w_{\rm ext}(\vx_E(t))$ reduces to $U_{\rm ext}
(t,\vx_E(t))$ in the Newtonian limit. The more accurate version of this
transformation will be discussed below.

Let us also note that the BCRS, GCRS and the transformation between
them have been discussed by \citet{Klioner:Soffel:2000} in the framework
of the PPN formalism with parameters $\beta$ and $\gamma$. For the
limit of General Relativity $\beta=\gamma=1$ all the formulas given in
that publication become equal to those derived in the framework of the
new IAU Resolutions that refer solely to Einstein's theory of gravity.

\subsection{Potential coefficients}
\label{Section-ML-SL}

\subsubsection{General post-Newtonian multipole moments}

For many problems it is advantageous to present the local gravitational
potentials of the Earth as multipole series that usually converge
everywhere outside the Earth. To this end one has to introduce a
certain set of multipole moments or {\it potential coefficients} for
the Earth. A certain set of potential coefficients, called
Blanchet-Damour (BD) moments
\citep{Blanchet:Damour:1989,Damour:Soffel:Xu:1991} defined to  first
post-Newtonian order has especially attractive features. Moreover, by
using such Blanchet-Damour moments we get a very simple form of the
multipole expansion of the post-Newtonian potentials (these expansions
have almost Newtonian form). Basically two sets of BD moments occur in
the formalism: mass multipole moments and spin multipole moments.
Theoretically these moments can be derived from the distribution of
mass and matter currents inside the body but for an observer they
simply present parameters which can be directly estimated from
observations.

Expressed in terms of symmetric and trace-free Cartesian tensors the BD
moments are denoted by ${\cal M}_L$ and ${\cal S}_L$. Here, $L$ is a
multi-index of $l$ different indices all taking the values $1,2,3$,
i.e., $L = i_1 i_2 \dots i_l$ and every index $i = 1,2,3$. Explicit
expressions for ${\cal M}_L$ and ${\cal S}_L$ as integrals over the
Earth can be found in, e.g.,
\citet{Blanchet:Damour:1989,Damour:Soffel:Xu:1991}

\begin{eqnarray}
{\cal M}_L(T) &\equiv& \int d^3X \, \hat X^L \Sigma
+ {1 \over 2(2l+3)c^2} {d^2 \over dT^2}
\left[ \int d^3X \, \hat X^L \vX^2 \Sigma \right] \nonumber \\
&& \ \ \ \ \ \ \ \ \
- {4(2l+1) \over (l+1) (2l+3) c^2}
{d \over dT}
\left[ \int \, d^3X \, \hat X^{aL} \Sigma^a \right]\, ,
 \quad l \ge 0 \, ,
\end{eqnarray}

\begin{eqnarray}
\label{spin-mom}
{\cal S}_L(T) &\equiv& \int \, d^3X \epsilon^{ab<c_l}
\hat X^{L-1>a}
\Sigma^b \, , \quad l \ge 1 \, ,
\end{eqnarray}

\noindent
where the integrations extend over the body under consideration and

\begin{equation}\label{Sigma-Sigmai}
\Sigma(T,\ve{X}) = {1\over c^2}\,\left({\cal T}^{00} + {\cal T}^{ss}\right) ,
\qquad
\Sigma^a(T,\ve{X}) = {1\over c}\,{\cal T}^{0a} \, .
\end{equation}

\noindent
Here ${\cal T^{\mu\nu}} = {\cal T}^{\mu\nu}(T,X^a)$ are the components
of the energy-momentum tensor in the GCRS. Both the caret and the sharp
brackets indicate the symmetric and trace-free (STF) part of the object
or of the indices enclosed by the brackets (see, e.g.
\citet{Damour:Soffel:Xu:1991} p. 3277 for the explicit definition of
the STF part of an object). Some basic information on the operations
with the STF objects can be found, e.g. in
\citet{Blanchet:Damour:1989,Damour:Soffel:Xu:1992}.

For practical applications, however, their explicit form  will not be
needed, since these quantities are parameters characterizing the
gravitational field of the corresponding body  which are fitted to
observations. The set ${\cal M}_L$ is equivalent to a set of potential
coefficients $C_{lm}$ and $S_{lm}$ that appear in a much more familiar
spherical harmonic
expansion of $W_E$. The first non-vanishing spin moment (the
spin dipole) of a body agrees with its spin vector (total intrinsic angular
momentum). The multipole expansion of $W_E$ and $W^a_E$ reads

\begin{eqnarray}\label{BD_expansion-W}
W_E&=&G\sum_{l=0}^\infty {(-1)^l\over l!}
\left[ {\cal M}_L\,\partial_L {1\over |\ve{X}|}+
{1\over 2c^2}\,\ddot {\cal M}_L
\,\partial_L |\ve{X}|\right]
+{4\over c^2}\Lambda_{,T}+\OO4,
\end{eqnarray}

\begin{eqnarray}\label{BD_expansion-Wa}
W^a_E&=&-G\sum_{l=1}^\infty {(-1)^l\over l!}
\left[
\dot{\cal M}_{aL-1} \partial_{L-1} {1\over |\ve{X}|}+
{l\over l+1} \epsilon_{abc} {\cal S}_{cL-1}
\partial_{bL-1} {1\over |\ve{X}|}
\right]
-\Lambda_{,a}+\OO2,
\end{eqnarray}

\noindent
where

\begin{eqnarray}\label{Lambda}
\Lambda&=&G \sum_{l=0}^\infty
{(-1)^l\over (l+1)!}\,{2l+1\over 2l+3}
\,{\cal P}_L\,\partial_L {1\over |\ve{X}|},
\end{eqnarray}

\begin{eqnarray}\label{PL}
{\cal P}_L=\int_V \Sigma^a \, \hat X^{aL}\,d^3X.
\end{eqnarray}

\noindent
Here the dot stands for $\partial/\partial T$ and $\partial_L$  for
$\partial^l/ \partial x^{i_1} \dots \partial x^{i_l}$. Also the comma
denotes partial differentiation, $\Lambda_{,T} \equiv
\partial \Lambda/\partial T$ and $\Lambda_{,a} \equiv
\partial \Lambda/\partial X^a$.

The gauge function $\Lambda$ does not enter the post-Newtonian
equations of motion. The latter contains only the BD multipole moments
${\cal M}_L$ and ${\cal S}_L$. The only place where the function
$\Lambda$ should be accounted for is in the transformation between the
various time scales. However, these gauge terms are of order $1/c^4$ in
the metric tensor so for the problem of clock rates they are
basically of second post-Newtonian order.
These terms are much less than $10^{-18}$ in
the geocentric metric tensor and will be neglected. For that reason the
$\Lambda$-terms are not mentioned in Resolution B1.4.

\subsubsection{Approximate expansion of the scalar gravitational potential}

A spherical harmonic expansion of $W_E$ equivalent to
(\ref{BD_expansion-W}) without the $\Lambda$ term reads $(R = \vert \vX
\vert)$

\begin{eqnarray}\label{BD-WE-sphe}
W_{E}(T,\vX)
&=&
{G M_{E} \over R}
\Bigl[ 1 + \sum_{l = 2}^\infty \sum_{m=0}^{+l}
\left( {R_{E} \over R} \right)^l P_{lm}(\cos \theta)
(\C_{lm}(T,R) \cos m \phi
+ \S_{lm}(T,R) \sin m \phi ) \Bigr] \nonumber \\
&& + \cO(c^{-4})
\end{eqnarray}

\noindent
with

\begin{eqnarray}
\C_{lm}^E(T,R) &=& C_{lm}^E(T) - {1 \over 2(2l-1)}
{R^2 \over c^2} {d^2 \over dT^2} C_{lm}^E(T)\\
\S_{lm}^E(T,R) &=& S_{lm}^E(T) - {1 \over 2(2l-1)}
{R^2 \over c^2} {d^2 \over dT^2} S_{lm}^E(T)\, .
\end{eqnarray}

\noindent
Let us stress that as stated in Resolution B1.4 $C_{lm}^E(T)$ and
$S_{lm}^E(T)$ refer to the GCRS coordinates and are related with
approximately constant potential coefficents in a terrestrial system
that is rotating with the Earth (i.e. those from an Earth's model) by
time-dependent transformations. For a rigid  axially symmetric body
rotating about its symmetry axis with angular velocity $\Omega_E$ the
second time derivative terms will vanish. Let us estimate these terms
for the Earth. From the order of magnitude of the $l=m=2$ terms in the
reference system corotating with the Earth one finds $C_{22}^E$ and
$S_{22}^E$ of order $10^{-6}$. The expected order of magnitude of the
second time derivative terms is $(\Omega_E\,R_E/c)^2 \simeq 10^{-12}$
times smaller than the corresponding 'Newtonian terms' from $C_{22}^E$
or $S_{22}^E$. The Newtonian terms lead to contributions in $G_{00}$ of
order $10^{-15}$ and hence the second time
derivative terms to contributions of order $10^{-27}$.
This is about nine orders of magnitude less than the $2 W^2/c^4$
term in $G_{00}$ which is of order $10^{-18}$. For that reason these
second time derivative terms in the Earth's metric can safely be
neglected at present. They are not mentioned in Resolution B1.4.

\subsubsection{Approximate expansion of the vector gravitational potential}

Let us now come to the gravitomagnetic vector potential of the Earth,
$W_E^a$. As can be seen from (\ref{BD_expansion-Wa}) this potential is
determined by the set of spin moments and the first time derivative of
the mass moments. As already mentioned, to characterize the
gravitational field outside of some matter distribution in GRT {\it
two} independent sets of multipole moments have to be used that in
principle should be determined from observational  data. So far the
spin moments of some astronomical body have not been studied and more
work is needed here. Formally the spin moments of the Earth are given
by expression (\ref{spin-mom}) above. Since for the post-Newtonian
metric we need these spin moments only to Newtonian order we might
proceed with a simple Newtonian model of a rigidly rotating  Earth with

\begin{displaymath}
\vSigma = \Sigma\, (\vOmega \times \vX) \, ,
\end{displaymath}

\noindent
where $\Sigma$ is the gravitational mass-energy density in the GCRS and
$\vOmega$ is the angular velocity of rotation that at this place has to
be defined only to Newtonian order. Under this assumption all spin
moments are proportional to the angular velocity and one might define a
set of Cartesian tensors $C_{Ld}$ such that

\begin{equation}
{\cal S}_L = C_{Ld} \Omega^d \, .
\end{equation}

\noindent
These tensors $C_{Ld}$ obey the following Newtonian relations

\begin{equation}
C_{Ld} = - M_{Ld} + {l+1 \over 2l+1 }
\delta_{d<a_l} N_{L-1>}  \, ,
\end{equation}

\noindent
where

\begin{equation}
M_L \equiv \int_E \Sigma \hat X^L \, d^3X \, , \qquad
N_L \equiv \int_E \Sigma \vX^2 \hat X^L \, d^3X \, .
\end{equation}

\noindent
Note that $C_{Ld}$ is symmetric and trace-free only in the first $L$
indices. Moreover, for the Newtonian mass moments $M_L$ one has

\begin{equation}
M_L = - C_{<L>} \, .
\end{equation}

\noindent
For a homogeneous ($\Sigma = \,$const.) and spherical Earth with radius
$R_E$ one finds for $l = 1$ the usual expression for the moment of
inertia tensor

\begin{displaymath}
C_{ab} = \delta_{ab}
\left( {2 \over 5} M R_E^2 \right)
\end{displaymath}

\noindent
that yields the total intrinsic angular momentum (spin) vector of the
Earth according to ${\cal S}_a=C_{ab}\,\Omega^b$. For a spherically
symmetric and mass centered Earth all  mass
moments $M_L$ with  $l \ge 1$ vanish and also all quantities $N_L$ with
$l > 0$. Hence in such a simple model only the spin vector is different
from zero and all higher spin moments vanish. For that reason we also
considered a rigidly rotating homogeneous oblate spheroid with equatorial
radius $A$ and polar radius $C$. For such a model all even spin
moments vanish since they are proportional to $C_L$ with odd $l$.
On the contrary, odd spin moments proportional to $C_L$ with even $l$
are nonzero.
For the spin dipole the usual result $C_{XX} = C_{YY} =
M(A^2 + C^2)/5$ and $C_{ZZ} = 2 M A^2/5$ for the moment of inertia
tensor is found. By means of computer algebra all components $C_L$ can
be found for any value of $l$. Let $\eta = (4 M A^4/525) \epsilon^2$
with $\epsilon^2 = (A^2 - C^2)/A^2 \simeq 2 f$, where $f$ is the usual
flattening. Assuming $\Omega^d = (0,0,\Omega)$ we found all
non-vanishing $l = 3$ terms, up to symmetries and terms of order $f^2$:
$S_{XXZ} = S_{YYZ} = 3 \eta\, \Omega$ and $S_{ZZZ} = - 6 \eta\,\Omega$.
This implies that the metric term resulting from the spin octupole of
the Earth near the Earth's surface is about $10^4$ times smaller than
the one from the spin dipole. In the following the contributions of
higher spin moments will be neglected.

Besides the spin moments the first time derivative of the mass moments
contribute to the gravitomagnetic field of  the Earth. For $l = 0$ we
encounter a $\dot {\cal M}_a$ term that vanishes if the post-Newtonian
center of mass condition ${\cal M}_a = 0$ is imposed. The next term is
given by
$\dot\cM_{ab}$ that is of order $\vert C^E_{22} \vert \, M R_E^2
\Omega$ and would vanish for an axially symmetric rigid body rotating
about its symmetry axis as well as the time derivative of all higher
mass moments. For the Earth the $\dot\cM_{ab}$ term is smaller than the
spin term (which is of order $ 2 M R_E^2 \Omega/5$) by a factor
determined by $C^E_{22} \simeq 1.6 \times 10^{-6}$ and hence negligible.
On the other hand, the vector potential $W_E^a(T, \vX)$ is employed
only in the calculations
of small relativistic effects (e.g. Lense-Thirring effects, higher-order
relativistic effects in the time transformations).
This implies that the expansion (\ref{BD_expansion-Wa}) for $W_E^a(T, \vX)$
can be truncated to the approximate expression

\begin{equation}\label{Wa-spin-only}
W_E^a(T, \vX) =  -{G\over 2} {(\vX \times \vS_{E})^a \over R^3} \,  ,
\end{equation}

\noindent
where $\vS_{E}$ is a vector with components ${\cal S}_a$.
This expression can be found in many standard textbooks on GRT
\citep{Weinberg:1972,Will:1993} and is usually related with
Lense-Thirring effects resulting from the Earth's rotational motion.

The reason why to characterize $W^a_E$ by the spin vector and not by the
angular velocity vector of the Earth is a conceptual one since usually it is
advantageous to characterize the gravitational field of the Earth in
the outside region by multipole moments. To get $W^a_E$ the Earth's
spin vector is needed only to Newtonian order and can be taken from current
precession-nutation models. Although one might use Newtonian concepts to
relate the gravitomagnetic field of the Earth with some Earth's angular
velocity, we prefer to employ the well-defined concept of multipole
moments here which are independent of any theoretical assumptions on
the rotational motion of the Earth.

\subsection{The barycentric metric in the mass monopole approximation}
\label{Section-mass-monopoles}

In the gravitational N-body problem the potential coefficents of a body
$A$ are defined in its corresponding local reference system (analogous
to the GCRS for the Earth). For many applications it is sufficient to
keep only the mass monopoles of the solar system  bodies, i.e., to put

\begin{equation}\label{ML=0}
\cM_L = 0 \quad {\rm for} \quad l \ge 1,
\qquad \cS_L = 0 \quad {\rm for} \quad l \ge 1
\end{equation}

\noindent
for all bodies and to keep the masses only, i.e., each body $A$ is
characterized by the value for its post-Newtonian mass ${\cal M}_A$ (we also put
$P_L = 0$). In the following we will use the symbol $M_A$ instead of ${\cal M}_A$
to be compatible with the text of the IAU 2000 Resolutions.

From the transformation rules for the metric potentials
(\ref{wE-WE}), expansions (\ref{BD_expansion-W})--(\ref{BD_expansion-Wa})
and formula (\ref{wsum})
one derives the metric in the barycentric coordinate system in the form
(\ref{bary_metric}) with

\begin{equation}
w = w_0 - {1 \over c^2} \Delta \,
\end{equation}

\noindent
where

\begin{equation}
\label{w_0}
w_0(t,\vx) \equiv \sum_A {G M_A \over r_A}
\end{equation}

\noindent
and

\begin{equation}
\Delta(t,\vx) = \sum_A \Delta_A(t,\vx)
\end{equation}

\noindent
with ($\ve{r}_{BA}=\ve{x}_B-\ve{x}_A$ and $\ve{a}_A=d\ve{v}_A/dt$)

\begin{eqnarray}
\label{delta-A}
\Delta_A(t,\vx) = && {G M_A \over r_A}
\left( - {3 \over 2} v_A^2 + \sum_{B \not= A}
{G M_B \over r_{BA} } \right)
 - {1 \over 2} G M_A \, r_{A,tt} \nonumber\\
= && {G M_A \over r_A}
\left[
- 2 v_A^2 + \sum_{B \not= A}
{G M_B \over r_{BA}}
+ {1 \over 2}
\left( {(r_A^k v_A^k)^2 \over r_A^2 } + r_A^k a_A^k \right)
\right]\, . \label{deltab}
\end{eqnarray}

\noindent
Furthermore, in our approximation

\begin{equation}
w^i(t,\vx) = \sum_A {G M_A \over r_A} v_A^i \, .
\end{equation}

\noindent
Note, that we have chosen the minus sign in front of $\Delta$ to have a
plus sign in the $c^{-4}$ part of $g_{00}$ (see Resolution B1.5). Note
furthermore, that the post-Newtonian Einstein-Infeld-Hoffmann equations
of motion for a system of mass monopoles, that form the basis of modern
solar system ephemerides, can be derived from that form of the
barycentric metric \citep[for details see][]{Damour:Soffel:Xu:1991}.
Thus, the barycentric mass monopole metric given above is already in
use for the description of the solar system dynamics.

One improvement of this simple mass monopole model is to consider the
spin dipoles of the various bodies as well (that is, to consider also
$\ve{S}_A$ to be non-zero). Actually Resolution B1.5 is based upon
such a mass monopole spin dipole model where modifications from the
simple mass monopole model are indicated explicitly.

\section {Time and frequency applications in the solar system}
\label{Section-time-frequency}

For practical applications concerning time and frequency measurements
in the solar system  it is necessary to consider a conventional model
for the realization of time coordinates and time transformations. This
model should be chosen so that i) its accuracy is significantly better
than the expected performance of clocks and time transfer techniques,
ii) it is consistent with the general framework of Section 3 and iii)
it may readily be used with existing astrometric quantities, e.g. solar
systems ephemerides.

Regarding item (i), we may derive reasonable accuracy limits for such a
model in a straightforward way. At present the best accuracies
are reached by Cs-fountain clocks operating
at less than two parts in $10^{15}$ in
fractional frequency \citep{Lemonde:et:al:2001,Weyers:et:al:2001}.
Their frequency stability
for time spans up to a few days
characterized by a standard Allan deviation is
of order $\sigma_y(\tau) = 4\times 10^{-14}\tau^{-1/2}$, for an
integration time $\tau$ in seconds.  In the near future, high accuracy
laser cooled Rb clocks \citep{Bize:et:al:1999} and space-borne Cs clocks
\citep{Lemonde:et:al:2001} are expected to reach accuracies of a few parts
in $10^{17}$ in fractional frequency and stabilities of order
$\sigma_y(\tau) = 1\times 10^{-14} \tau^{-1/2}$. The uncertainty in the
time transformations should induce errors that are always lower than
the expected performance of these future clocks. Including a factor 2
as safety margin, we therefore conclude that time coordinates and time
transformations should be realized with an uncertainty not larger than
$5 \times 10^{-18}$ in rate or, for quasi-periodic terms, not larger
than $5 \times 10^{-18}$ in rate  and 0.2 ps in amplitude.

For the spatial domain of validity of the transformations, we note that
projects like SORT (Solar Orbit Relativity Test) plan to fly highly
accurate  clocks to within 0.25 AU of the Sun, which is therefore the
lower limit for the distance to the barycenter that we will consider.
In the geocentric system we will consider locations from the Earth's
surface up to geostationary orbits ($|\ve{X}|<50\,000$~km).

To comply with item (ii), we render the  developments following the
general framework outlined in Section 3 and we show (iii) how the time
transformations, e.g., $TCB-TCG$, may be performed with the existing
astrometric quantities and tools.

\subsection {Barycentric reference system}
\label{Section-time-BCRS}

Let us write the barycentric metric potential $w(t,\vx)$ in the form

\begin{equation}
\label{wparts}
w = w_0 + w_L - {1 \over c^2} \Delta \, ,
\end{equation}

\noindent
where $w_L$ contains contributions from higher potential coefficients
with $l \ge 1$ and can be determined from equation
(\ref{BD_expansion-W}) and the transformation rules of the metric
potentials. Evaluating the $\Delta_A$ terms from Resolution B1.5
(equation (\ref{delta-A}) plus spin terms) for all bodies of the solar
system, we find that in the metric tensor $|\Delta_A(t,{\bf x})|/c^4$
may reach at most a few parts in $10^{17}$ in the vicinity of Jupiter
and about $10^{-17}$ close to the Earth. Presently, however, for all
planets except the Earth, the magnitude of $\Delta_A(t,{\bf x})/c^4$ in
the vicinity of the planet is smaller than the uncertainty in $w_0/c^2$
or $w_L/c^2$ originating from the uncertainties in its mass multipole
moments so that it is practically not needed to account for these
terms. Nevertheless, when new astrometric observations allow to derive
the moments with adequate uncertainty, it will be necessary to do so.
In any case, for the vicinity of a given body $A$, only the effect of
$\Delta_A(t,{\bf x})$ is needed in practice, i.e., the effect of
$\sum_{B\not=A} \Delta_B(t,{\bf x})$ is smaller than our accuracy
specifications. For the comparison of proper time of a clock in the
vicinity of the Earth with that of other clocks in the solar system or
with $TCB$ it may thus be needed to account for $\Delta_E(t,{\bf
x})/c^4$.

From (\ref{bary_metric}) and (\ref{wparts})  the transformation between
proper time of some observer  and $TCB$ may be derived within
our accuracy limit

\begin{equation}
{\rm d}\tau / {\rm d}TCB = 1-{1\over c^2}\left( w_0+w_L+{v^2\over 2} \right)
+ {1\over c^4} \left(-{1 \over 8}v^4 - {3 \over 2}v^2 w_0 + 4v^i w^i +
{1 \over 2} w_0^2 + \Delta \right) \, ,
\label{tauTCB}
\end{equation}

\noindent
where $v^i$ is the BCRS coordinate velocity of the observer. Similarly,
the transformation between $TCB$ and $TCG$ in the immediate vicinity of
the Earth, accurate to the limits as specified  above can be derived
from the general post-Newtonian $TCB - TCG$ transformation from
Resolution B1.3 as

\begin{eqnarray}
\label{TCB-TCG_T-F}
TCB - TCG &=& c^{-2}\left[\int^{t}_{t_{0}}\left({v_{E}^{2} \over 2} +
{w}_{0,{\rm ext}}({\bf x}_ {E}) \right)dt + v^i_E r^i_E \right]
\nonumber \\
  &-& c^{-4}[\int^t_{t_0}\left(-{1 \over 8}v^4_E - {3 \over 2}v_E^2
{w}_{0,{\rm ext}}({\bf x}_E) + 4v_E^i {w}^i_{\rm ext}({\bf x}_E)
+ {1 \over 2} {w}_{0,{\rm ext}}^2({\bf x}_E) \right)dt
\nonumber \\
 && - (3 {w}_{0,{\rm ext}} ({\bf x}_E) + v_E^2/2) v_E^i r^i_E] \, ,
\end{eqnarray}

\noindent
where $t$ is $TCB$. Here ${w}_{0,{\rm ext}}$ is defined by (\ref{w_0}) with
summation over all solar system bodies expect the Earth.
Note that $t_0$ was not explicitly defined in
Resolution B1.5 (2000). It is the origin of $TCB$ and $TCG$, defined in
Resolution A4 (1991) (see Section \ref{Section-IAU-1991}). The external
contributions to $w_L$ and $\Delta$ are beyond our accuracy limit and
can be neglected here.

This equation is composed of terms evaluated at the geocenter (the two
integrals) and of position dependent terms linear in $|\vr_E|$,
terms with higher powers of $|\vr_E|$ having been found to be
negligible. The integrals may be computed from existing planetary
ephemerides \citep{Fukushima:1995,Irwin:Fukushima:1999}. Since, in
general, the planetary ephemerides are expressed in terms of a time
argument $T_{\rm eph} = (1 - L_B) \, TCB+T^0_{\rm eph}$
\citep{Standish:1998,Irwin:Fukushima:1999}, the first integral will be
computed as

\begin{equation}
\int^{t}_{t_{0}}\left({v_{E}^{2} \over 2} + {w}_{0,{\rm ext}}({\bf x}_ {E})
\right)dt = \left[ \int^{T_{\rm eph}}_{T_{{\rm eph}_0}}\left({v_{E}^{2} \over 2}
+ {w}_{0,{\rm ext}}({\bf x}_ {E}) \right)dT_{\rm eph}
 \right] / (1 - L_B)
\, .
\end{equation}

Terms in the second integral of (\ref{TCB-TCG_T-F}) are secular and
quasi periodic. They amount to $\sim 1.1 \times 10^{-16}$ in rate
$({\rm d}TCB/{\rm d}TCG)$ and primarily a yearly term of $\sim 30$ ps
in amplitude (i.e. corresponding to periodic rate variations of
amplitude $\sim 6\times 10^{-18}$). Position-dependent terms in
$c^{-4}$ are not negligible and reach, for example, an amplitude of 0.4
ps ($\sim 3\times 10^{-17}$ in rate) in geostationary orbit.

\subsection {Geocentric reference system}
\label{Section-time-GCRS}

Evaluating the contributions of the different terms in the metric
tensor of the GCRS given in Resolution B1.3 to the ${\rm d}\tau/{\rm
d}(TCG)$ transformation on the Earth's surface and up to geostationary
orbit we find that terms of order $c^{-2}$ reach 7 parts in $10^{10}$
while the contributions from $W^2$ and $W^a$ do not exceed 5 parts in
$10^{19}$. Also, the terms from $W_{\iner}$ in $W$ remain below $2\times
10^{-20}$. Therefore, the terms given in the IAU 1991 framework with
the metric of the form (\ref{iau_metric}) are sufficient for time and
frequency applications in the GCRS in the region $\vert\vX\vert <
50\,000\,$km for present and foreseeable future clock accuracies. Note,
some care needs to be taken when evaluating the potential $W$ at the
location of the clock which is not trivial when accuracies of order
$10^{-18}$ are required
\citep{Klioner:1992,Petit:Wolf:1994,Wolf:Petit:1995}.

Presently, the time scale of reference for all practical issues on
Earth is Terrestrial Time ($TT$) or one of the scales realizing it and
differing by some time offset (e.g., $TAI$, $UTC$, $GPS$ time). $TT$
was defined in IAU Resolution A4 (1991) as: "a time scale differing
from the Geocentric Coordinate Time $TCG$ by a constant rate, the unit
of measurement of $TT$ being chosen so that it agrees with the SI
second on the geoid". According to the transformation between proper
and coordinate time, this constant rate is given by ${\rm d}(TT) / {\rm
d}(TCG) = 1-U_g/c^2 = 1 - L_G$, where $U_g$ is the gravity
(gravitational + rotational) potential on the geoid (this notation is
used instead of the usual '$W_0$' to avoid confusion with GCRS
gravitational potential $W$ used throughout the paper).

Some shortcomings appeared in this definition of $TT$ when considering
accuracies below $10^{-17}$. First, the uncertainty in the
determination of $U_g$ is of order $1\ {\rm m}^2\,{\rm s}^{-2}$ or
slightly better \citep{Bursa:1995,Groten:1999}. Second, even if it is
expected that the uncertainty in $U_g$ improves with time the  surface
of the geoid is difficult to realize (so that it is difficult to
determine the potential difference between the geoid and the location
of a clock). Third, the geoid is, in principle, variable with time.
Therefore it was decided to desociate the definition of $TT$ from the
geoid while maintaining continuity with the previous definition. The
constant $L_G$ was turned into a defining constant with its value fixed
to $6.969290134 \times 10^{-10}$ (Resolution B1.9, see Appendix
\ref{Appendix-IAU-2000}). This removes the limitations mentioned above
when realizing $TT$ from clocks on board of terrestrial satellites
(such as in the GPS). In Table \ref{Table-1} we present numerical
values for the constants $L_C$, $L_G$ and $L_B$ relating the mean rates
of different relativistic timescales.

\begin{table}
\begin{tabular}{lrrr}
&IAU 1991&IAU 2000&IAU 2000\\
&s/s&s/s&ms/yr\\
\\
\hline\\
$L_C$&$1.480813\times 10^{-8}$&$1.48082686741 \times 10^{-8}$&467.313\\
$L_G$&$6.969291\times 10^{-10}$&$6.969290134\times10^{-10}$&21.993\\
$L_B\equiv L_C+L_G-L_C\,L_G$&$1.550505\times10^{-8}$&$1.55051976772\times10^{-8}$&489.307\\
\end{tabular}
\caption{The constants relating the mean rates of different
relativistic timescales. Both the values adopted by the IAU 1991
recommendations and the IAU 2000 resolutions are given. As an
illustration the IAU 2000 values are also given in milliseconds per
Julian year.
\label{Table-1}
}
\end{table}

\section{Final remarks}

The relativistic Resolutions of the IAU represent a post-Newtonian
framework allowing one to model any kind of astronomical observations
in a rigorous, self-consistent manner with accuracies that are
sufficient for the next decades. They replace the old IAU relativistic
framework that was insufficient for many reasons discussed above. These
new Resolutions, however, are not expected to  lead to dramatic
changes. In fact in many fields of application the models presently in
use are already fully compatible with the new IAU Resolutions and in
this sense the IAU Resolutions officially fix the status quo. Let us
give some examples for that.

The metric tensor of the BCRS allows one to derive the
Einstein-Infeld-Hoffman equations which have been used since the 70s to
construct the JPL numerical ephemerides of planetary motion
\citep{Newhall:Standish:Williams:1983}. The BCRS is the basic astrometric
reference system where concepts such as 'radial velocity' or 'proper motion'
are defined \citep{Lindegren:Dravins:2003}.
The metric tensors of both GCRS
and BCRS and the transformation between corresponding coordinates were
used to formulate the VLBI model employed by the IERS since 1992
\citep{IERS:1992,IERS:1996,IERS:2003}. The equations of motion of the
Earth's satellites recommended by the IERS
\citep{IERS:1992,IERS:1996,IERS:2003} are compatible with the new IAU
framework and can be derived from the given metric tensor of the GCRS.

The models used for costructing the HIPPARCOS catalog make it clear
that this catalog represents a materialization of the BCRS. The full
power of the new IAU theoretical framework will be needed to construct
the model for astrometric positional observations with an accuracy of 1
microarcsecond which will be necessary for future astrometric missions.
Work in this direction has already started \citep{Klioner:2003}.

It is obvious that this explanatory supplement presents only a first
step to show how the new IAU Resolutions concerning relativity should
be employed in practice. Much more work will be  necessary to reach
that goal.


\acknowledgements
The anonymous referee is thanked for his valuable suggestions to
improve the text and to make it more readable.


\newpage

\appendix

\section{IAU Resolutions concerning
Relativity Adopted at the 24th General Assembly}
\label{Appendix-IAU-2000}

\subsection*{Resolution B1.3: {Definition of Barycentric Celestial
Reference System and Geocentric Celestial Reference System}}

\bigskip\bn
\noindent The XXVIth International Astronomical Union General Assembly\hfil\break

\noindent \underline{\bf Considering} \hfil\break

{\parskip 0pc
\begin{itemize}

\item[1.] that the Resolution A4 of the XXIst General Assembly (1991)
has defined a system of space-time coordinates for (a) the solar system
(now called the Barycentric Celestial Reference System, (BCRS)) and (b)
the Earth (now called the Geocentric Celestial Reference System
(GCRS)), within the framework of General Relativity,

\item[2.] the desire to write the metric tensors both in the BCRS and
in the GCRS in a compact and self-consistent form, and

\item[3.] the fact that considerable work in General Relativity has
been done using the harmonic gauge that was found to be a useful and
simplifying gauge for many kinds of applications,

\end{itemize}
}

\noindent \underline{\bf Recommends} \hfil\break

{\parskip 0pc
\begin{itemize}

\item[1.] the choice of harmonic coordinates both for the barycentric
and for the  geocentric reference systems,

\item[2.] writing the time-time component and the space-space component
of the barycentric metric $g_{\mu\nu}$ with barycentric coordinates
$(t, {\bf x})$ ($t$ = Barycentric Coordinate Time ($TCB$)) with a single
scalar potential $w(t,{\bf x})$ that generalizes the Newtonian
potential, and the space-time component with a vector potential
$w^i(t, {\bf x})$; as a boundary condition it is assumed that these two
potentials vanish far from the solar system,

explicitly,

\hfil\break

\begin{eqnarray*}
g_{00} &=& -1 + {2w \over c^2} - {2w^2 \over c^4}, \\
g_{0i} &=& -{4 \over c^3} w^i, \\
g_{ij} &=& \delta_{ij}\left(1 + {2 \over c^2}w \right),
\end{eqnarray*}

\noindent
with

\begin{eqnarray*}
w(t,{\bf x}) &=& G \int d^3 x' {\sigma(t,{\bf x'}) \over
 |{\bf x} - {\bf x'}|}
+ {1 \over 2c^2} G {\partial^2 \over \partial t^2} \int d^3 x' \sigma(t,{\bf x'})
|{\bf x} - {\bf x'}| \\
w^i(t,{\bf x}) &=&
G \int d^3 x' {\sigma^i(t,{\bf x'}) \over |{\bf x} - {\bf x'}|}.
\end{eqnarray*}

\noindent
Here, $\sigma$ and $\sigma^i$ are the gravitational mass and current
densities, respectively.

\smallskip
\item[3.] writing the geocentric metric tensor $G_{\alpha\beta}$ with
geocentric coordinates $(T, {\bf X})$ ($T$ = Geocentric Coordinate Time
($TCG$)) in the same form as the barycentric one but with potentials
$W(T, {\bf X})$ and $W^a(T, {\bf X})$; these geocentric potentials
should be split into two parts --- potentials $W_E$ and $W_E^a$ arising
from the gravitational action of the Earth and external parts $W_{\rm
ext}$ and $W^a_{\rm ext}$ due to tidal and inertial effects; the
external parts of the metric potentials are assumed to vanish at the
geocenter and admit an expansion into positive powers of ${\bf X}$,

explicitly,

\begin{eqnarray*}
G_{00} &=& -1 + {2W \over c^2} - {2W^2 \over c^4}, \\
G_{0a} &=& -{4 \over c^3} W^a, \\
G_{ab} &=& \delta_{ab}\left(1 + {2 \over c^2}W \right).
\end{eqnarray*}

\noindent
The potentials $W$ and $W^a$ should be split according to

\begin{eqnarray*}
W(T,{\bf X}) &=& W_E(T,{\bf X}) + W_{\rm ext}(T,{\bf X}), \\
W^a(T,{\bf X}) &=& W_E^a(T,{\bf X}) + W_{\rm ext}^a(T,{\bf X}).
\end{eqnarray*}

\item[] The Earth's potentials $W_E$ and $W_E^a$ are defined in the
same way as $w$ and $w^a$ but with quantities calculated in the
GCRS with integrals taken over the whole Earth.

\smallskip
\item[4.] using, if accuracy requires, the full post-Newtonian
coordinate transformation between the BCRS and the GCRS as induced by
the form of the corresponding metric tensors,

\item[]explicitly, for the kinematically non-rotating GCRS ($T$ = $TCG$,
$t$ = $TCB$, $r_E^i \equiv x^i - x^i_E(t)$ and a summation from 1 to 3
over equal indices is implied),

\begin{eqnarray*}
T &=& t - {1 \over c^2} \biggl[ A(t) + v_E^i r_E^i\biggr] + {1 \over c^4}
\biggl[B(t) + B^i(t)r_E^i + B^{ij}(t)r_E^i r_E^j + C(t,{\bf x})\biggr ]
+ O(c^{-5}), \\
X^a &=& \delta_{ai} \biggl[ r_E^i + {1 \over c^2}
\biggl({1 \over 2} v_E^i  v_E^j
r_E^j + w_{\rm ext}({\bf x_E})r_E^i + r_E^i a_E^j r_E^j - {1 \over 2} a_E^i r_E^2
\biggr) \biggr] + O(c^{-4}),
\end{eqnarray*}

\noindent
where

\begin{eqnarray*}
{d \over dt}A(t) &=& {1 \over 2} v_E^2 + w_{\rm ext}({\bf x_E}), \\
{d \over dt}B(t) &=& -{1 \over 8}
v_E^4 - {3 \over 2} v_E^2 w_{\rm ext}({\bf x_E})
+ 4v_E^iw_{\rm ext}^i({\bf x_E}) + {1 \over 2}w_{\rm ext}^2({\bf x_E}), \\
B^i(t) &=& -{1 \over 2}v_E^2 v_E^i + 4 w_{\rm ext}^i({\bf
x_E}) -
3v_E^iw_{\rm ext}
({\bf x_E}), \\
B^{ij}(t) &=&
-v_E^i\delta_{aj}Q^a + 2{\partial \over \partial x^j}w_{\rm ext}^i
({\bf x_E}) - v_E^i{\partial \over \partial x^j} w_{\rm ext}({\bf x_E})
+  {1 \over 2}\delta^{ij}\dot{w}_{\rm ext}({\bf x_E}), \\
C(t,{\bf x}) &=& -{1 \over 10}r_E^2(\dot{a}_E^i r_E^i).
\end{eqnarray*}

\item[] Here $x_E^i$, $v_E^i$, and $a_E^i$ are the barycentric
position, velocity and acceleration vectors of the Earth, the dot
stands for the total derivative with respect to $t$, and

$$
Q^a = \delta_{ai} \left[{\partial \over \partial x_i} w_{\rm ext}({\bf x_E})
- a_E^i\right]\, .
$$

\noindent
The external potentials, $w_{\rm ext}$ and $w_{\rm ext}^i$, are given by

$$
w_{\rm ext}=\sum_{A\not= E}
w_A, \quad w_{\rm ext}^i = \sum_{A\not= E} w_A^i\, ,
$$

\item[] where $E$ stands for the Earth and $w_A$ and $w_A^i$ are
determined by the expressions for $w$ and $w^i$ with integrals taken
over body $A$ only.

\end{itemize}
}

\vskip1truecm

\noindent \underline{\bf Notes} \hfil\break

{\it

\noindent
It is to be understood that these expressions for $w$ and $w^i$ give
$g_{00}$ correct up to $\cO(c^{-5})$, $g_{0i}$ up to $\cO(c^{-5})$, and
$g_{ij}$ up to $\cO(c^{-4})$. The densities $\sigma$ and $\sigma^i$ are
determined by the components of the energy momentum tensor of the
matter composing the solar system bodies as given in the references.
Accuracies for $G_{\alpha\beta}$ in terms of $c^{-n}$ correspond to those of
$g_{\mu\nu}$.

\noindent
The external potentials $W_{\rm ext}$ and $W_{\rm ext}^a$ can be
written in the form

\begin{eqnarray*}
W_{\rm ext} &=& W_{\rm tidal} + W_{\rm iner}, \\
W_{\rm ext}^a &=& W_{\rm tidal}^a + W_{\rm iner}^a \, .
\end{eqnarray*}

\noindent
$W_{\rm tidal}$ generalizes the Newtonian expression for the tidal
potential. Post-Newtonian expressions for $W_{\rm tidal}$ and $W_{\rm
tidal}^a$ can be found in the references. The potentials $W_{\rm
iner}$, $W_{\rm iner}^a$ are inertial contributions that are linear in
$X^a$. The former is determined mainly by the coupling of the Earth's
nonsphericity to the external potential. In the kinematically
non-rotating Geocentric Celestial Reference System, W$_{\rm iner}^a$
describes the Coriolis force induced mainly by geodesic precession.

\noindent
Finally, the local gravitational potentials $W_E$ and $W_E^a$ of the
Earth are related to the barycentric gravitational potentials $w_E$ and
$w_E^i$ by

\begin{eqnarray*}
W_E(T,{\bf X}) &=& w_E(t,{\bf x})\left(1 + {2 \over c^2}v_E^2 \right) -
{4 \over c^2} v_E^i w_E^i(t,{\bf x}) + O(c^{-4}), \\
W_E^a(T,{\bf X}) &=&
\delta_{ai}(w_E^i(t,{\bf x}) - v_E^i w_E(t,{\bf x}))
+ O(c^{-2})\, .
\end{eqnarray*}

}

\vskip1truecm
\noindent \underline{\it References} \hfil\break

\noindent Brumberg, V.A., Kopeikin, S.M., 1989, {\it Nuovo Cimento}, B {\bf 103},
63. \hfil\break
Brumberg, V.A., 1991, {\it Essential Relativistic Celestial Mechanics}, Hilger,
Bristol. \hfil\break
Damour, T., Soffel, M., Xu, C., {\it Phys. Rev. D}, {\bf 43}, 3273 (1991);
{\bf 45}, 1017 (1992); {\bf 47}, 3124 (1993); {\bf 49}, 618 (1994). \hfil\break
Klioner, S. A., Voinov, A.V., 1993, {\it Phys Rev. D}, {\bf 48}, 1451. \hfil\break
Kopeikin, S.M., 1988, {\it Celest. Mech.}, {\bf 44}, 87. \hfil\break

\subsection*{Resolution B1.4: {Post-Newtonian Potential Coefficients}}

\smallskip
\noindent The XXVIth International Astronomical Union General Assembly,
\hfil\break

\noindent \underline{\bf Considering} \hfil\break

{\parskip 0pc
\begin{itemize}

\item[1.] that for many applications in the fields of celestial
mechanics and astrometry a suitable parametrization of the metric
potentials (or multipole moments) outside the massive solar-system
bodies in the form of expansions in terms of potential coefficients are
extremely useful, and

\item[2.] that physically meaningful post-Newtonian potential
coefficients can be derived from the literature,

\end{itemize}
}

\noindent \underline{\bf Recommends} \hfil\break

{\parskip 0pc
\begin{itemize}

\item[1.] expansion of the post-Newtonian potential of the Earth in the
Geocentric Celestial Reference System (GCRS) outside the Earth in the
form

$$
W_E(T,{\bf X}) = {GM_E \over R} \Biggl[1 +
\sum_{l=2}^\infty
\sum_{m=0}^{+l}
\biggl({R_E \over R} \biggr )^l P_{lm}(\cos\theta)\bigl(C_{lm}^E(T)\cos m\phi
+ S_{lm}^E(T)\sin m\phi\bigr ) \Biggr]\, .
$$

\item[] Here $C_{lm}^E$ and $S_{lm}^E$ are, to sufficient accuracy,
equivalent to the post-Newtonian multipole moments introduced by Damour
{\it et al.} (Damour {\it et al.}, {\it Phys. Rev. D}, {\bf 43}, 3273,
1991). $\theta$ and $\phi$ are the polar angles corresponding to the
spatial coordinates X$^a$ of the GCRS and $R=|X|$, and

\item[2.] expression of the vector potential outside the Earth, leading
to the well-known Lense-Thirring effect, in terms of the Earth's total
angular momentum vector ${\bf S_E}$ in the form

$$
W_E^a(T,{\bf X}) = -{G \over 2}{({\bf X}\times {\bf
S_E})^a
\over R^3}\, .
$$

\end{itemize}
}

\subsection*{Resolution B1.5: {Extended relativistic framework for
time transformations and realization of coordinate times in the solar
system}}

\bigskip
\noindent
The XXVIth International Astronomical Union General Assembly,
\hfil\break

\noindent \underline{\bf Considering} \hfil\break

{\parskip 0pc
\begin{itemize}

\item[1.] that the Resolution A4 of the XXIst General Assembly(1991)
has defined systems of space-time coordinates for the solar system
(Barycentric Reference System) and for the Earth (Geocentric Reference
System), within the framework of General Relativity,

\item[2.] that Resolution B1.3 entitled ``Definition of Barycentric
Celestial Reference System and Geocentric Celestial Reference System"
has renamed these systems the Barycentric Celestial Reference System
(BCRS) and the Geocentric Celestial Reference System (GCRS),
respectively, and has specified a general framework for expressing
their metric tensor and defining coordinate transformations  at the
first post-Newtonian level,

\item[3.] that, based on the anticipated performance of atomic clocks,
future time and frequency measurements will require practical
application of this framework in the BCRS, and

\item[4.] that theoretical work requiring such expansions has already
been performed,

\end{itemize}
}

\noindent
\underline{\bf Recommends} \hfil\break

\noindent
that for applications that concern time transformations and realization
of coordinate times within the solar system, Resolution B1.3 be applied
as follows:

{\parskip 0pc
\begin{itemize}

\item[1.] the metric tensor be expressed as

\begin{eqnarray*}
g_{00} &=& -\biggl ( 1 - {2 \over c^2} (w_0(t,{\bf x}) + w_L(t,{\bf x})) +
{2 \over c^4}(w_0^2(t,{\bf x}) + \Delta(t,{\bf x}))\biggr ) \\
g_{0i} &=& -{4 \over c^3} w^i(t,{\bf x}), \\
g_{ij} &=& \biggl ( 1 + {2w_0(t,{\bf x}) \over c^2} \biggr
)\delta_{ij},
\end{eqnarray*}

\item[] where ($t \equiv$ Barycentric Coordinate Time ($TCB$), ${\bf x}$)
are the barycentric coordinates, $w_0=G\sum_A M_A/r_A$ with the
summation carried out over all solar system bodies $A$, ${\bf r}_A =
{\bf x} - {\bf x}_A, {\bf x}_A$ are the coordinates of the center of
mass of body $A$, $r_A$ = $|{\bf r_A}|$, and where $w_L$ contains the
expansion in terms of multipole moments [see their definition in the
Resolution B1.4 entitled ``Post-Newtonian Potential Coefficients"]
required for each body. The vector potential $w^i(t,{\bf x}) =\sum_A
w_A^i(t,{\bf x})$ and the function $\Delta(t,{\bf x}) = \sum_A \Delta_A
(t,{\bf x})$ are given in note 2.

\item[2.] the relation between $TCB$ and Geocentric Coordinate Time
($TCG$) can be expressed to sufficient accuracy by

\begin{eqnarray}
&&TCB-TCG = c^{-2}\Biggl[\int_{t_0}^t{\biggl({v_E^2 \over 2} + w_{0,{\rm
ext}}({\bf x}_E)\biggr)} dt + v_E^i r_E^i\Biggr ]
\nonumber\\
&&-c^{-4}\Biggl[\int_{t_0}^t\biggl(-{1 \over 8} v_E^4 - {3 \over
2} v_E^2 w_{0,{\rm ext}}({\bf x}_E) + 4v_E^iw_{\rm ext}^i({\bf x}_E) +
{1 \over 2}w_{0,{\rm  ext}}^2({\bf x}_E)\biggr)dt -\biggl(3w_{0,{\rm
ext}}({\bf x}_E) + {v_E^2 \over 2}\biggr)v_E^i r_E^i \Biggr ],
\nonumber
\end{eqnarray}

\noindent where $v_E$ is the barycentric velocity of the Earth and where the index
ext refers to summation over all bodies except the Earth.
\end{itemize}
}

\noindent \underline{\bf Notes} \hfil\break

{\it

\noindent
1. This formulation will provide an uncertainty not larger than $5
\times 10^{-18}$ in rate and, for quasi-periodic terms, not larger than
$5 \times 10^{-18}$ in rate amplitude and 0.2 ps in phase amplitude,
for locations farther than a few solar radii from the Sun. The same
uncertainty also applies to the transformation between $TCB$ and $TCG$ for
locations within $50\,000\,$km of the Earth. Uncertainties in the
values of astronomical quantities may induce larger errors in the
formulas.

\hfil\break
\noindent
2. Within the above mentioned uncertainties, it is sufficient to
express the  vector potential $w_A^i(t,{\bf x})$ of body A as

$$
w_A^i(t,{\bf x}) = G \Biggl[{-({\bf r}_A \times {\bf S}_A)^i \over 2r_A^3} + {M_A v_A^i \over r_A}
\Biggr ],
$$

\noindent
where ${\bf S}_A$ is the total angular momentum of body $A$ and $v_A^i$
is the barycentric coordinate velocity of body $A$. As for the function
$\Delta_A(t,{\bf x})$ it is sufficient to express it as

$$
\Delta_A(t,{\bf x}) = {GM_A \over r_A} \Biggl [-2v_a^2 +
\sum_{B\not=A}{GM_B \over
 r_{BA}}
+ {1 \over 2}\biggl({(r_A^kv_A^k)^2 \over r_A^2} + r_A^k a_A^k \biggr ) \Biggr ]
+ {2Gv_A^k({\bf r}_A \times {\bf S}_A)^k \over r_A^3},
$$

\noindent
where $r_{BA}=|{\bf x}_B-{\bf x}_A|$ and $a_A^k$ is the barycentric
coordinate acceleration of body $A$. In these formulas, the terms in
${\bf S}_A$ are needed only for Jupiter ($S \approx 6.9 \times
10^{38}\,$m$^2$s$^{-1}$kg) and Saturn ($S \approx 1.4 \times
10^{38}\,$m$^2$s$^{-1}$kg), in the immediate vicinity of these planets.

\hfil\break
\noindent
3. Because the present recommendation provides an extension of the IAU
1991 recommendations valid at the full first post-Newtonian level, the
constants $L_C$ and $L_B$ that were introduced in the IAU 1991
recommendations should be defined as $<TCG/TCB> = 1 - L_C$ and
$<TT/TCB> = 1 - L_B$, where $TT$ refers to Terrestrial Time and $<>$
refers to a sufficiently long average taken at the geocenter. The most
recent estimate of $L_C$ is (Irwin, A. and Fukushima, T., 1999, {\it Astron.
Astroph.}, {\bf 348}, 642--652)

$$
L_C = 1.48082686741 \times 10^{-8} \pm 2 \times 10^{-17}.
$$

From Resolution B1.9 on ``Redefinition of Terrestrial Time $TT$", one
infers $L_B = 1.55051976772 \, $ $ \times 10^{-8} \pm 2 \times
10^{-17}$ by using the relation $1-L_B=(1-L_C)(1-L_G)$. $L_G$ is
defined in Resolution B1.9.

Because no unambiguous definition may be provided for $L_B$ and $L_C$,
these constants should not be used in formulating time transformations
when it would require knowing their value with an uncertainty of order
$1 \times 10^{-16}$ or less.
 \hfil\break

\noindent
4. If $TCB - TCG$ is computed using planetary ephemerides which are
expressed in terms of a time argument (noted $T_{\rm eph}$) which is
close to Barycentric Dynamical Time ($TDB$), rather than in terms of
$TCB$, the first integral in Recommendation 2 above may be computed as

$$
\int_{t_0}^t \biggl({v_E^2 \over 2} + w_{0,\rm ext}({\bf x}_E) \biggr) dt =
\Biggl[ \int_{T_{\rm eph_0}}^{T_{\rm eph}} \biggl({v_E^2 \over 2} +
w_{0,\rm ext}({\bf x}_E)
\biggr ) dt \Biggr ] / (1-L_B).
$$

}

\subsection*{Resolution B1.9: Re-definition of Terrestrial Time $TT$}

\noindent
The XXVIth International Astronomical Union General Assembly,
\hfil\break

\noindent \underline{\bf Considering} \hfil\break

{\parskip 0pc
\begin{itemize}

\item[1.] that IAU Resolution A4 (1991) has defined Terrestrial Time
($TT$) in its Recommendation 4, and

\item[2.] that the intricacy and temporal changes inherent to the
definition and realization of the geoid are a source of uncertainty in
the definition and realization of $TT$, which may become, in the near
future, the dominant source of uncertainty in realizing $TT$ from atomic
clocks,

\end{itemize}
}

\noindent \underline{\bf Recommends} \hfil\break

\noindent
that $TT$ be a time scale differing from $TCG$ by a constant rate:
$dTT/dTCG= 1-L_G$, where $L_G = 6.969290134\times10^{-10}$ is a defining
constant,

{\it

\noindent \underline{\bf Note}
\hfil\break

\noindent
$L_G$ was defined by the IAU Resolution A4 (1991) in its Recommendation
4 as equal to $U_G/c^2$ where $U_G$ is the geopotential at the geoid.
$L_G$ is now used as a defining constant.

}

\newpage

\section{Comparison of the IAU metric with versions
given in the literature}
\label{Appendix-IAU-PPN}

In this Appendix we will compare the metric (\ref{bary_metric}) with
well-known results from the literature. Only for this purpose we will
consider the material composing the various bodies of the system  to
behave like an ideal fluid (for the IAU 2000 Resolutions this is not
assumed). In the ideal fluid case the energy-momentum tensor can be
written in the form

\begin{eqnarray}
\label{Tmn}
T^{00} &=& \rho c^2
\left( 1 + {1 \over c^2} \left[ \Pi + \vv^2 + 2U \right]
\right) + \OO2,\nonumber \\
T^{0i} &=& \rho c v^i + \OO1,\label{ener_fluid} \\
T^{ij} &=& \rho v^i v^j + p \delta_{ij} + \OO2\, . \nonumber
\end{eqnarray}

\noindent
Here, $\rho$ denotes the rest-mass density, $p$ is the pressure, $\Pi$
is the specific internal energy \citep[e.g.,][]{Will:1993},
$v^i(t,\vx)$ is the velocity of the corresponding material element and

\begin{equation}
U(t,\vx) \equiv
G \int {\rho(t,\vx') \over
\vert \vx - \vx' \vert } \, d^3x' \, .
\end{equation}

\noindent
From (\ref{sigma}) and (\ref{Tmn}) we derive

\begin{eqnarray}
\sigma &=& \rho
\left( 1 + {1 \over c^2}
\left[ \Pi + 2 \vv^2 + 2 U \right] \right) + 3 {p \over c^2} +
\OO4 \, ,\nonumber \\
\sigma^i &=& \rho v^i + \OO2  \, .
\end{eqnarray}

\noindent
Introducing the metric potentials

\begin{eqnarray}
\Phi_1 \equiv \int {\rho' v'{}^2 \over \vert\vx - \vx'\vert }
\, d^3x' \, ,\qquad
\Phi_2 \equiv \int {\rho' U' \over \vert\vx - \vx'\vert }
\, d^3x' \, ,\\
\Phi_3 \equiv \int {\rho' \Pi' \over \vert\vx - \vx'\vert }
\, d^3x' \, ,\qquad
\Phi_4 \equiv \int { p' \over \vert\vx - \vx'\vert }
\, d^3x' \, ,
\end{eqnarray}

\noindent
we obtain from (\ref{fielda}) and (\ref{fieldb})

\begin{eqnarray}
w &=& U + 2 \Phi_1 + 2 \Phi_2 + \Phi_3 + 3 \Phi_4
- {1 \over 2 c^2} \chi_{,tt} + \OO4 \, , \nonumber \\
w^i &=& V_i+\OO2 \, ,
\end{eqnarray}

\noindent
with

\begin{equation}
V_i \equiv \int {\rho' v_i' \over \vert \vx - \vx' \vert }
\, d^3x' \, ,\qquad
\chi \equiv - G \int \rho' \vert \vx - \vx' \vert \, d^3x' \, ,
\end{equation}

\noindent
and the comma denotes partial differentiation as in

\beqn
\chi_{,tt} \equiv {\partial^2 \chi \over \partial t^2} \, .
\eeqn

\noindent
The post-Newtonian metric (in harmonic gauge) can then be written as

\begin{eqnarray}
g_{00} &=& - 1 + {1 \over c^2}
\left[ 2 U + 4 \Phi_1 + 4 \Phi_2 + 2 \Phi_3 + 6
\Phi_4 \right] - {2 \over c^4} U^2 + {1 \over c^4} \chi_{,tt} \, ,
\nonumber \\
g_{0i} &=& - {4 \over c^3} V_i \, ,\\
g_{ij} &=& \delta_{ij}
\left( 1 + {2U \over c^2} \right) \, .\nonumber
\end{eqnarray}

\noindent
To compare, e.g., with the metric in \citep{Will:1993} we transform
from harmonic coordinates, used in the present paper and recommended by
the IAU, to standard post-Newtonian (SPN) coordinates used by several
authors including Will. This is achieved by a gauge transformation of
the form (e.g., Eq. (3.12) of \citet{Damour:Soffel:Xu:1991}),

\begin{equation}
w_{\rm SPN} = w - {1 \over c^2} \lambda_{,t}\,  ,
\qquad
w^i_{\rm SPN} = w^i + {1 \over 4} \lambda_{,i}
\end{equation}

\noindent
with

\begin{equation}
\lambda = - {1 \over 2} \chi_{,t} \, .
\end{equation}

\noindent
This implies that the $\chi$-term disappears from $w$ and hence from
$g_{00}$ when standard post-Newtonian coordinates are employed, but the
$g_{0i}$ term gets affected by this transformation. Using the relation

\beqn
\chi_{,ti} = V_i - W_i
\eeqn

\noindent
with

\beqn
W_i \equiv \int {\rho' \,[ \vv' \cdot (\vx - \vx')]\,
(x^i - x'^i) \over
\vert \vx - \vx' \vert^3 } \, d^3x'
\eeqn

\noindent
one verifies that the metric induced by the potentials (\ref{fielda})
and (\ref{fieldb}) agrees in General Relativity with (5.28) of
\citep{Will:1993}.


\end{document}